\documentclass[preprintnumbers,showpacs,amsmath,amssymb,floatfix,9pt,prd,twocolumn,
superscriptaddress,nofootinbib]{revtex4}
\usepackage{graphicx}
\usepackage{latexsym}
\usepackage{epsfig}
\usepackage{amssymb}
\usepackage[utf8]{inputenc}
\begin{document}

\title{Charged compact star in $f(R,T)$ gravity in Tolman-Kuchowicz spacetime}

\author{Pramit Rej\footnote{Corresponding author}}
\email{pramitrej@gmail.com
 } \affiliation{Department of
Mathematics, Sarat Centenary College, Dhaniakhali, Hooghly, West Bengal 712 302, India}

\author{Piyali Bhar}
\email{piyalibhar90@gmail.com
 } \affiliation{Department of
Mathematics,Government General Degree College, Singur, Hooghly, West Bengal 712 409,
India}

\author{Megan Govender}
\email{megandhreng@dut.ac.za
 } \affiliation{Department of Mathematics, Faculty of Applied Sciences,
Durban University of Technology, Durban, South Africa}

\begin{abstract}

In this current study, our main focus is to model a specific charged compact star SAX J 1808.4-3658 (M = 0.88 $M_{\odot}$,\, R = 8.9 km)  within the realm of $f(R,\,T)$ modified gravity theory using the metric potentials proposed by Tolman-Kuchowicz (Tolman, Phys Rev 55:364, 1939; Kuchowicz  Acta Phys Pol 33:541, 1968) and the interior spacetime is matched to the exterior Reissner-Nordstr\"{o}m line element at the surface of the star. Tolman-Kuchowicz metric potentials provide a singularity-free solution which satisfies the stability criteria. Here we have used the simplified phenomenological MIT bag model equation of state (EoS) to solve Einstein-Maxwell field equations where the density profile ($\rho$) is related to the radial pressure ($p_r$) as $p_r(r) = (\rho - 4B_g)/3$. Further, to derive the values of unknown constants $a,\, b,\, B,\, C$ and the bag constant $B_g$, we match our interior space-time to the exterior Reissner-Nordstr\"{o}m line element at the surface of stellar system. In addition to this, to check the physical validity and stability of our suggested model, we evaluate some important properties such as effective energy density, effective pressures, radial and transverse sound velocities, relativistic adiabatic index, all energy conditions, compactness factor and surface redshift. It is depicted from our current study that all our derived results lie within the physically accepted regime which provides the viability of our present model in the context of $f(R,\,T)$ modified gravity.
\end{abstract}


\maketitle

\maketitle

\section{Introduction}


Einstein's general relativity (GR) has continued to withstand the test of time in its predictions of physical phenomena within the realms of astrophysics and cosmology. From the classical predictions of the precession of Mercury's orbit and the deflection of starlight by a massive gravitating body to present day detection of gravitational waves and observations of black holes GR has triumphed. Early attempts seeking solutions of the Einstein field equations which describe stellar objects were crude and for most part unrealistic. The first exact solution of the Einstein field equations describing a self-gravitating sphere was obtained by Karl Schwarzschild. The so-called interior Schwarzschild solution which described a constant density sphere suffered various pathologies, the most notable being that the propagation speed for any signals within the fluid sphere was noncausal \cite{Sch1916}. A survey of exact solutions appearing in the literature describing stellar objects by Delgaty and Lake \cite{lake1} revealed that only a small subset of solutions meet the rigorous tests for physical viability, regularity and stability of fluid spheres.

The search for more realistic stellar models within GR required researchers to connect the macroscopic properties of stars determined through observations to the microphysics. A new era of stellar modeling was born which went beyond the mathematical excursion of the Einstein field equations where ad-hoc assumptions were made just to generate a toy model. Standard approaches which included assumptions on the metric function, density profiles, pressure profiles, anisotropy parameter and even the matter content which allowed for the system of equations to be integrated gave way to well-motivated techniques intrinsically connected to physics which include an equation of state (EoS), mass profiles linked to surface redshift and compactness of typical stellar structures. The linear EoS which links the radial pressure to the energy density has been generalised to include the microphysics (at least on a phenomenological level) via the so-called MIT bag model. The departure from pressure isotropy makes the modeling of stellar objects mathematically tractable. Imposing a barotropic EoS of the form $p_r = p_r(\rho)$ reduces the problem of finding an exact model of the Einstein field equations to a single generating function. The imposition of an EoS has richly rewarded researchers with a handle on understanding complicated microphysics on a macroscale\cite{lin2,lin1,q1,cg1}. The colour-flavour locked (CFL) EoS describing strange quark matter was shown to mimic the linear EoS. The CFL EoS has also been used to investigate the surface tension of stars which placed tighter restrictions on the bag constant as well as tangential pressure associated with the model. It has been recently demonstrated that the pressure isotropy condition is unstable. A self-gravitating sphere in quasi-static equilibrium will evolve into a regime in which the radial and transverse pressures are unequal. A shear-free fluid sphere will evolve into a shear-like epoch due to contributions from the Weyl tensor, density inhomogeneities and/or the presence of dissipative fluxes \cite{89a}.

Though general relativity (GR) has been used successfully to predict various phenomena that
Newtonian gravitation fails to explain, there are still many unresolved issues. After the
discovery of the accelerated expansion of the Universe, several extended theories of gravity
have been developed, like $f(R)$ gravity, $f(R,\,T)$ gravity, $f(T)$ gravity, $f(G,\,T)$ gravity, EGB gravity etc. In our present paper, we will consider $f(R,\,T)$ gravity, which has got immense interest in recent years. In fact, $f(R,\,T)$ gravity was phenomenologically introduced by Harko et al.\cite{harko2}. In this connection, we want to mention that recently the anisotropic charged and uncharged models in modified $f(R,\,T )$ theory gravity have been studied in \cite{frt1,frt2,frt3,frt4,frt5}. Pretel et al. \cite{pretel} examined the static structure configurations and radial stability of compact
stars within the context of $f(R,\,T)$ gravity. Mahanta \cite{mahanta} constructed Locally Rotationally Symmetric Bianchi type I (LRSBI) cosmological models in the $f(R,\,T)$ theory of gravity when the source of gravitation is the bulk viscous fluid. The models were constructed for $f(R,\,T)=R+2f(T)$ and $f(R,\,T)=f_1(R)+f_2(T)$. Singh et al. \cite{singh1} have studied flat Friedmann-Lemaitre-Robertson-Walker (FLRW) model with modified Chaplygin gas (MCG) in $f(R,\,T)$ gravity with particle creation. The cosmological reconstruction of the Little Rip model in $f(R,\,T)$ gravity was investigated by Houndjo et al. \cite{hound} which perfectly reproduces the present stage of the universe, characterized by the $\Lambda$CDM model, without singularity at future finite-time (without the Big Rip). The problem of violation of causality in $f(R,\,T)$ modified gravity was proposed by Santos and Ferst \cite{sf}. Static spherically symmetric wormholes in $f(R,\,T)$ gravity was proposed by Zubair et al. \cite{zubair}. Azmat and Zubair \cite{az} have adopted gravitational decoupling by minimal geometric deformation (MGD) approach and have developed an anisotropic version of well-known Tolman VII isotropic solution in the framework of $f(R, T)$ gravity.

Charged compact objects have been studied for about a century within the context of classical GR. The Einstein-Maxwell field equations can be interpreted as an anisotropic system in which the electric field intensity mimics the anisotropy factor. These toy models have helped us understanding the (in)stability of static fluid spheres in the presence of charge. There have been various mechanisms put forward to account for a significant residual charge in stars. Several researchers have discovered the solution of the Einstein-Maxwell equations to describe the model of strange quark stars,
charged black hole and other astrophysical compact objects \cite{c1,c2,c3,c4,c5,c6}. Kiczek and Rogatko \cite{c7} studied the properties of ultra-compact spherically symmetric dark matter sector star objects, being the solution of Einstein equations with two U(1)-gauge fields, first one is the ordinary Maxwell field, while the second one is auxiliary gauge field pertains to the hidden
sector, and mimics the properties of dark matter. Arba\~{n}il and Malheiro \cite{arb} studied the hydrostatic equilibrium and the stability against radial perturbation of charged strange quark stars composed of a charged perfect fluid. To construct the model the author considered the perfect fluid
follows the MIT bag model equation of state and the radial charge distribution follows a power-law. Negreiros and collaborators \cite{n1,n2} modeled compact objects considering spheres composed of strange matter that
follows the MIT bag model equation of state (EoS) and
a Gaussian distribution of the electric charge on the surface
of the star. In this paper the authors estimated that
the electric charge that causes significant impact on the
structure of the strange stars produces a surface electric
field of the order $E \sim 10^{22}$ [V/m].

We have organized the paper as follows: In Sec.~\ref{sec2} we briefly summarize $f(R,\,T)$ gravity and we
present the corresponding relativistic equations within the framework of $f(R,\,T )= R+2\gamma T$
model in presence of electric field. Sec.~\ref{sec4} deals with the solution of the field equations by choosing Tolman-Kuchowicz {\em ansatz}. In next section we have described about various physical properties of our present model analytically as well as graphically. The stability of the model has been studied under various forces in Sec.~\ref{sec8}. Finally in Sec.~\ref{dis}, our conclusions are summarized.

\section{Basic field Equations in $f(R,\,T)$ gravity with charge}\label{sec2}
The general formulation of
Einstein-Hilbert (EH) action in General Relativity is expressed by,
\begin{eqnarray}
S_{EH}&=&\frac{1}{16 \pi}\int d^4x \sqrt{-g}R
\end{eqnarray}
The above expression of the action in $f(R,T)$ theory of gravity in the presence of charge is modified as
\cite{h58},
\begin{eqnarray}\label{action}
S&=&\frac{1}{16 \pi}\int  f(R,T)\sqrt{-g} d^4 x + \int \mathcal{L}_m\sqrt{-g} d^4 x \nonumber\\
&&+\int \mathcal{L}_e\sqrt{-g} d^4 x,
\end{eqnarray}
where $g = det(g_{\mu \nu}$), $f(R,\,T )$ represents the general function of Ricci scalar $R$ and trace $T$ of the energy-momentum tensor $T_{\mu \nu}$, $\mathcal{L}_m$ and $\mathcal{L}_e$ respectively denote the lagrangian matter density and Lagrangian for the electromagnetic field.\\
Let us assume a static spherically symmetric spacetime in curvature coordinates $(t,\,r,\,\theta,\,\phi)$ as,
\begin{eqnarray}\label{line}
ds^{2}&=&-e^{\nu}dt^{2}+e^{\lambda}dr^{2}+r^{2}\left[\sin^{2}\theta d\phi^{2}+d\theta^{^2}\right],
\end{eqnarray}
the metric potentials $\nu$ and $\lambda$ are purely depends on radial co-ordinate `r' and it ranges from $0$ to
$\infty$. The main aim of our present work is to obtain a physically reasonable and singularity free model of compact star and for this reason our present paper is developed by utilizing the following {\em ansatz}
\begin{eqnarray}\label{elambda}
e^{\lambda}= 1 + ar^2 + br^4, e^{\nu}&=&C^2 e^{B r^2},
\end{eqnarray}
where $a, b, B, C$ are constants. This metric potentials are well-known as Tolman-Kuchowicz {\em ansatz} \cite{tk1,tk2} and it was successfully used earlier by several authors to model the compact star both in the context of general relativity and modified gravity. In the background of Tolman-Kuchowicz spacetime, Javed et al.\cite{javed} obtained anisotropic spheres in $f(R,\,G)$ modified gravity, Majid and Sharif \cite{majid} obtained Quark Stars in Massive Brans-Dicke Gravity, Farasat Shamir and Fayyaz \cite{ff} obtained the model of charged compact star in $f(R)$ gravity, Naz and Shamir \cite{naz2} found the stellar model in $f(G)$ gravity and its charged version \cite{naz1}, Biswas et al. \cite{biswas} obtained anisotropic strange star with $f(R,\,T)$ gravity, Bhar et al. \cite{pb4} modelled compact star in EGB modified gravity.\\
It is worthly to mention that for asymptotically flat
spacetime both the metric potential $\nu(r)$ and $\lambda(r)$ tends to $0$ as $r~\rightarrow~\infty$ but in our present case this condition is not satisfied.
Here, we have taken the signature of the spacetime as $(-,\, +,\, +,\,+)$. Now the Einstein-Maxwell field equations
for obtaining the hydrostatic stellar structure of the charged
sphere in modified $f(R,\,T)$ gravity corresponding to action (\ref{action}) is given by,
\begin{eqnarray}\label{frt}
f_R R_{\mu \nu}-\frac{1}{2} g_{\mu \nu} f +(g_{\mu \nu }\Box-\nabla_{\mu}\nabla_{\nu})f_R &=& 8\pi (T_{\mu \nu}+E_{\mu \nu})\nonumber\\&&-f_T (T_{\mu \nu}+  \Theta_{\mu \nu}).\nonumber\\
\end{eqnarray}
Where, $f=f(R,T)$, $f_R(R,T)=\frac{\partial f(R,T)}{\partial R},~f_T(R,T)=\frac{\partial f(R,T)}{\partial T}$. $\nabla_{\nu}$ represents the covariant derivative associated with the Levi-Civita connection of $g_{\mu \nu}$, $\Theta_{\mu \nu}=g^{\alpha \beta}\frac{\delta T_{\alpha \beta}}{\delta g^{\mu \nu}}$ and
$\Box \equiv \frac{1}{\sqrt{-g}}\partial_{\mu}(\sqrt{-g}g^{\mu \nu}\partial_{\nu})$ represents the D'Alembert operator.\par
Landau and Lifshitz \cite{landau} defined the stress-energy tensor of matter as,
\begin{eqnarray}\label{tmu1}
T_{\mu \nu}&=&-\frac{2}{\sqrt{-g}}\frac{\delta \sqrt{-g}\mathcal{L}_m}{\delta \sqrt{g_{\mu \nu}}},
\end{eqnarray}
and its trace $T$ is defined by, $T=g^{\mu \nu}T_{\mu \nu}$. Now, if the Lagrangian density $\mathcal{L}_m$ depends only on $g_{\mu \nu}$, not on its derivatives, the above equation of $T_{\mu \nu}$ takes the following form :
\begin{eqnarray}
T_{\mu \nu}&=& g_{\mu \nu}\mathcal{L}_m-2\frac{\partial \mathcal{L}_m}{\partial g_{\mu \nu}},
\end{eqnarray}
Now the matter Lagrangian density could be a function of both density and pressure, $\mathcal{L}_m = \mathcal{L}_m (\rho,\, p)$, or it only becomes an arbitrary function of the density of the matter $\rho$ only, so that $\mathcal{L}_m = \mathcal{L}_m(\rho)$ \cite{har1}. For our present paper, we choose the matter Lagrangian as $\mathcal{L}_m=\rho$ and the expression of $\Theta_{\mu \nu}=-2T_{\mu \nu}-\rho g_{\mu\nu}.$\\
Let us assume that the underlying fluid distribution of our proposed model is anisotropic in nature and therefore $T_{\mu\nu}$ in
anisotropic fluid form having the components \begin{eqnarray}
T_{\mu \nu}= diag (-\rho,\,p_r,\,p_t,\,p_t),
 \end{eqnarray}
$\rho$ being the matter density, $p_r$ and $p_t$ are respectively the radial and transverse pressure in modified gravity. The electromagnetic energy-momentum tensor $E_{\mu \nu}$ has the following form :
\begin{eqnarray}
E_{\mu \nu}&=&\frac{1}{4 \pi}\left(F_{\mu}^{\alpha}F_{\nu \alpha}-\frac{1}{4}F^{\alpha \beta}F_{\alpha \beta} g_{\mu \nu}\right),
\end{eqnarray}
where, $F_{\mu \nu}$ is the antisymmetric
electromagnetic field strength tensor defined by
\begin{eqnarray}
F_{\mu \nu}&=&\frac{\partial A_{\nu}}{\partial x^{\mu}}-\frac{\partial A_{\mu}}{\partial x^{\nu}},
\end{eqnarray}
and it satisfies the Maxwell equations,
\begin{eqnarray}
F^{\mu \nu}_{;\nu}=\frac{1}{\sqrt{-g}}\frac{\partial}{\partial x^{\nu}}(\sqrt{-g}F^{\mu \nu})&=&-4\pi j^{\mu},\label{ta}\\
F_{\mu\nu;\lambda}+F_{\nu \lambda;\mu}+F_{\lambda \mu;\nu}&=&0.
\end{eqnarray}
where $A_{\nu}=(\phi(r),\,0,\,0,\,0)$ is the four-potential and  $j^{\mu}$ is the
four-current vector, defined by
\begin{eqnarray}
j^{\mu}&=&\frac{\rho_e}{\sqrt{g_{00}}}\frac{dx^{\mu}}{dx^0},
\end{eqnarray}
where $\rho_e$ denotes the proper charge density.
The expression for the electric field can be obtained from Eq.~(\ref{ta}) as follows,
\begin{eqnarray}
F^{01}&=&-e^{\frac{\lambda+\nu}{2}}\frac{q(r)}{r^2},
\end{eqnarray}
here $q(r)$ represents the net charge inside a sphere of radius `r' and it can be obtained as,
\begin{eqnarray}
q(r)&=& 4\pi \int_0^r \rho_e e^{\frac{\lambda}{2}} r^2 dr.
\end{eqnarray}
In order to discuss
the coupling effects of matter and curvature components in
$f (R,\,T )$ gravity, let us consider a
separable functional form given by
\begin{eqnarray}\label{e}
f(R,\,T)&=& R+2 \gamma T,
\end{eqnarray}
where $\gamma$ is some small positive constant. Harko et al. \cite{h58} proposed that for $\gamma~\rightarrow~0$, the Eq. (\ref{e}) produces the field equations in General Relativity. The term $2 \gamma T$ induces time-dependent coupling between curvature and
matter.\\
For the line element (\ref{line}), the field equations in modified gravity can be written as,
\begin{eqnarray}
8\pi\rho^{\text{eff}}+\frac{q^2}{r^4}&=&\frac{\lambda'}{r}e^{-\lambda}+\frac{1}{r^{2}}(1-e^{-\lambda}),\label{f1}\\
8 \pi p_r^{\text{eff}}-\frac{q^2}{r^4}&=& \frac{1}{r^{2}}(e^{-\lambda}-1)+\frac{\nu'}{r}e^{-\lambda},\label{f2} \\
8 \pi p_t^{\text{eff}}+\frac{q^2}{r^4}&=&\frac{1}{4}e^{-\lambda}\left[2\nu''+\nu'^2-\lambda'\nu'+\frac{2}{r}(\nu'-\lambda')\right].\nonumber\\ \label{f3}
\end{eqnarray}
The quantity $q(r)$ actually determines the electric field as,
\begin{eqnarray}
E(r)&=&\frac{q(r)}{r^2}.
\end{eqnarray}
where $\rho^{\text{eff}}$, $p_r^{\text{eff}}$ and $p_t^{\text{eff}}$ are respectively the density and pressures in Einstein Gravity where
\begin{eqnarray}
\rho^{\text{eff}}&=& \rho+\frac{\gamma}{8\pi}( \rho-p_r-2p_t),\label{r1}\\
p_r^{\text{eff}}&=& p_r+\frac{\gamma}{8\pi}(\rho+3p_r+2p_t),\label{r2}\\
p_t^{\text{eff}}&=& p_t+\frac{\gamma}{8\pi}(\rho+p_r+4p_t).\label{r3}
\end{eqnarray}
The prime denotes differentiation with respect to `r'. Our aim is to find the solution of the system (\ref{f1})-(\ref{f3}) which will fully specify the behavior of the interior of the stellar object. $\rho,\,p_r$ and $p_t$ respectively denote the matter density and pressures in modified gravity. \\
Now by the taking covariant divergence of (\ref{frt}), the divergence of the stress-energy tensor $T_{\mu \nu}$ can be obtained as \cite{h58,hi,farri},
\begin{eqnarray}\label{conservation}
\nabla^{\mu}T_{\mu \nu}&=&\frac{f_T(R,T)}{8\pi-f_T(R,T)}\Big[(T_{\mu \nu}+\Theta_{\mu \nu})\nabla^{\mu}\ln f_T(R,T)\nonumber\\&&+\nabla^{\mu}\Theta_{\mu \nu}-\frac{1}{2}g_{\mu \nu}\nabla^{\mu}T-\frac{8\pi}{f_T}\nabla^{\mu}E_{\mu \nu}\Big].
\end{eqnarray}
From eqn.(\ref{conservation}), we can check that $\nabla^{\mu}T_{\mu \nu}\neq 0$ if $f_T(R,T)\neq 0$. So the system will not be conserved like Einstein gravity. In the next section we are interested to find the solutions of the field equations in charged case in $f(R,\,T)$ modified theory of gravitation.
\section{Our proposed model in $f(R,\,T)$ modified gravity}\label{sec4}
Employing the expressions of the metric coefficients given in (\ref{elambda}), into the equations (\ref{f1})-(\ref{f3}), the following set of equations are obtained :
\begin{eqnarray}
8\pi \rho^{\text{eff}}+\frac{q^2}{r^4}&=&\frac{3 a + (a^2 + 5 b) r^2 + 2 a b r^4 + b^2 r^6}{{\Psi}^2},\label{s1}\\
8\pi p_r^{\text{eff}}-\frac{q^2}{r^4}&=& -\frac{a - 2 B + b r^2}{\Psi},\label{s2}\\
8 \pi p_t^{\text{eff}}+\frac{q^2}{r^4}&=&\frac{1}{{\Psi}^2}\Big[-a + 2 B + \big(
   B (a + B)-2b\big) r^2 +\nonumber\\&&a B^2 r^4 + b B^2 r^6\Big],\label{s3}
\end{eqnarray}
where, $\Psi$ is a function of `r' given by,
\begin{eqnarray*}
\Psi&=& (1 + a r^2 + b r^4).
\end{eqnarray*}
To describe strange quark matter, a very general approach is to use MIT bag model equation of state. A stellar object whose matter content
consists only of up, down and strange quarks, MIT bag model is the simplest equation
of state to study the equilibrium configuration of the model. This equation of state takes into account that
these quarks are massless and non-interacting quarks confined by a bag constant $B_g$. For the
anisotropic fluid studied, we consider that the energy density and the radial pressure of the
fluid are connected through the relation:
\begin{eqnarray}\label{s9}
\rho&=& 3p_r+4 B_g ,
\end{eqnarray}
Here, $B_g$ is defined as the bag constant. Mak
and Harko investigated the difference
between the bag constant and mass density of the perturbed
and non-perturbed QCD vacuum, the units of bag constant in MeVfm$^{-3}$ is
derived by Chodos et al. \cite{chodos}.\\
The equation (\ref{s9}) can be written as,
\begin{eqnarray}\label{s4}
p_r&=& \frac{1}{3} (\rho- 4 B_g).
\end{eqnarray}
Witten \cite{witten} proposed that the formation of strange matter can be classified into two ways: the
quark-hadron phase transition in the early universe and conversion of neutron stars into strange stars at ultra high densities. Farhi and Jaffe \cite{sj}
showed that, for massless and non-interacting quarks, the Witten conjecture is verified
for a bag constant approximately between the values $57$ MeV/fm$^3$ and $94$ MeV/fm$^3$. For our present paper we consider $B_g = 60$ MeV/fm$^3$.\\
With the help of (\ref{s4}), we solve the eqns. (\ref{s1})-(\ref{s3}) and obtain the expressions for matter density and pressures in Einstein gravity as follows:
\begin{eqnarray}
\rho^{\text{eff}}&=& {B_g} +\frac{ 3 (a + B + (2 b + a B) r^2 + b B r^4)}{16 \pi {\Psi}^2},\label{24}\\
p_r^{\text{eff}}&=& -{ B_g} +\frac{a + B + (2 b + a B) r^2 + b B r^4}{16 \pi {\Psi}^2},\label{25}\\
p_t^{\text{eff}}&=& {B_g} + \frac{1}{16 \pi {\Psi}^2} \Big[-5 a + 7 B + (-2 a^2 - 8 b
   \nonumber\\&& + 5 a B + 2 B^2) r^2 + \big(3 b B + 2 a (-2 b + B^2)\big) r^4
   \nonumber\\&& + 2 b (-b + B^2) r^6 \Big].\label{26}
\end{eqnarray}
The expression for the anisotropic factor in general relativity is given by,
\begin{eqnarray}
  \Delta^{\text{eff}} &=&p_t^{\text{eff}}-p_r^{\text{eff}} \nonumber\\
  &=& 2~{ B_g} +\frac{-2 a + 2 B - 4 b r^2 + a B r^2}{8 \pi {\Psi}^2}  \nonumber\\&&+
   \frac{-a + B + (-b + B^2) r^2}{8 \pi \Psi}.
\end{eqnarray}
\begin{figure}[htbp]
    \centering
        \includegraphics[scale=.45]{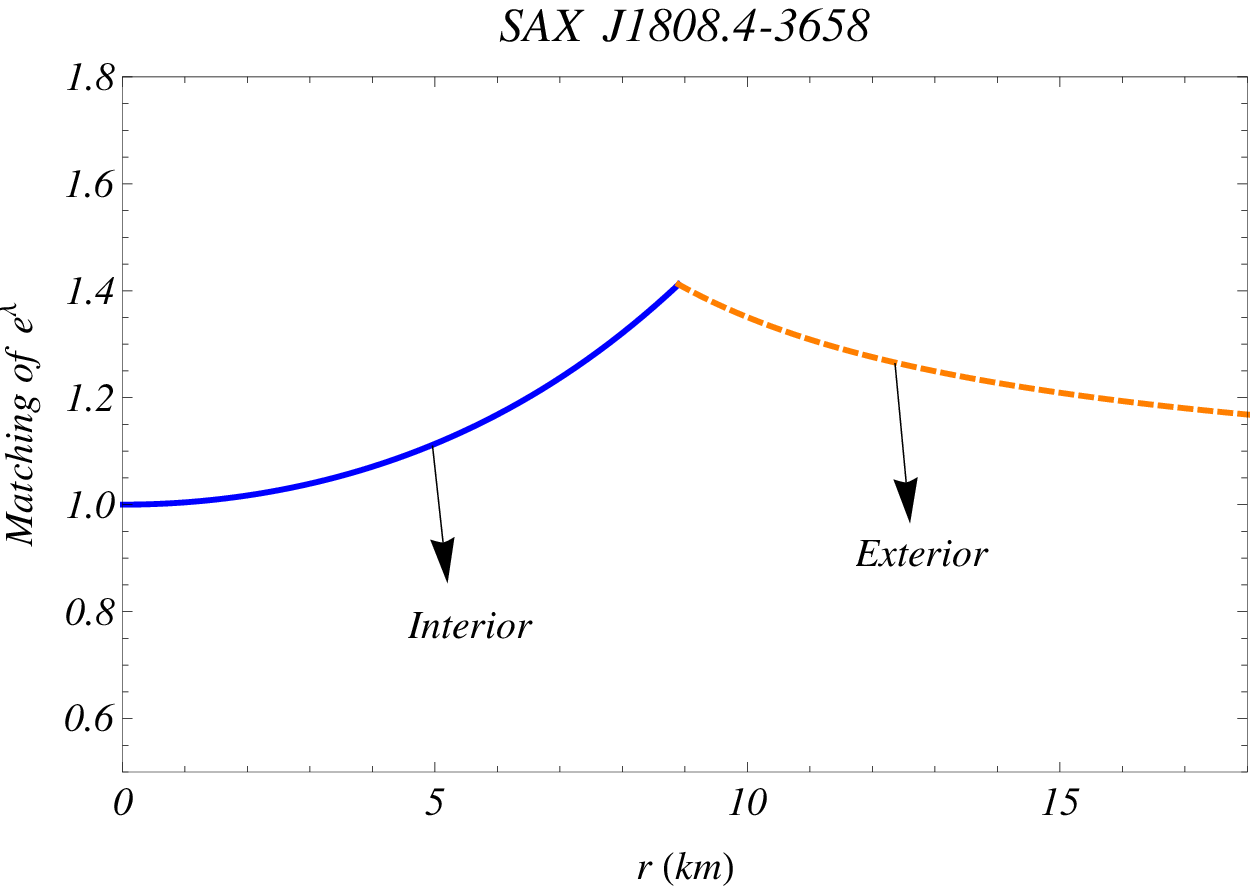}
        \includegraphics[scale=.45]{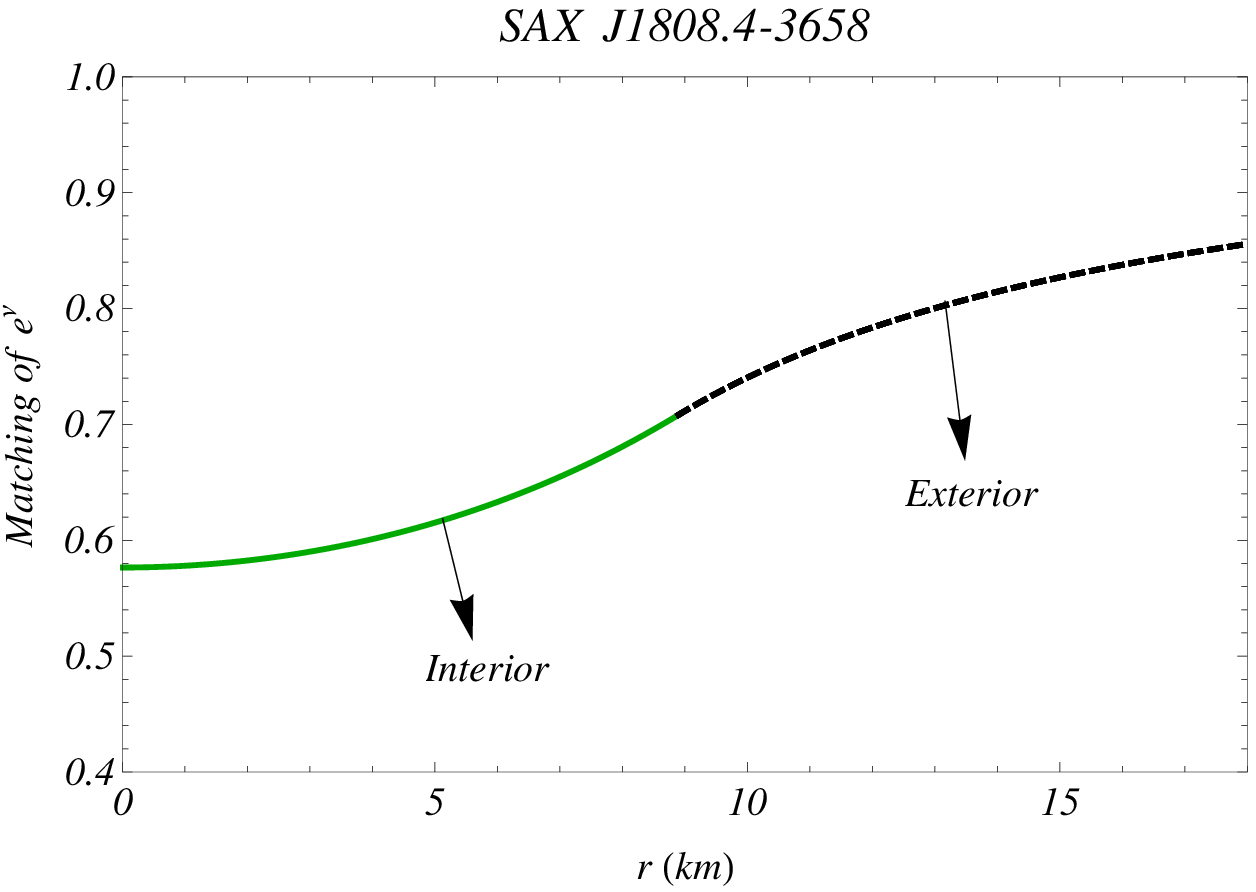}
       \caption{The matching condition of the metric potential $e^{\lambda}$ and $e^{\nu}$ are
shown against radius for the compact star SAX J $1808.4-3658$. \label{metric}}
\end{figure}
The electric field $E^2$ in modified gravity is obtained as,
\begin{eqnarray}\label{k4}
E^2&=& \frac{1}{3\gamma + 4 \pi}\bigg[-4~{ B_g} (\gamma + 2 \pi) (\gamma + 4 \pi)\nonumber\\&& +\frac{1}{\Psi}\Big\{2 a (\gamma + 2 \pi)  - 2 B (\gamma + 3 \pi)\nonumber\\&&+\Big(B^2 \gamma + 2 b (\gamma + 2 \pi)\Big) r^2\Big\}+\frac{1}{{\Psi}^2} \Big\{2 B \gamma \nonumber\\&& +2 b (\gamma + 2 \pi) r^2 + a (\gamma + 2 \pi + B \gamma r^2)\Big\}\bigg].
\end{eqnarray}
To obtain the expressions for $\rho,\,p_r$ and $p_t$ we solve the eqns.(\ref{r1})-(\ref{r3}) by using the expressions given in (\ref{24})-(\ref{26}). Solving those equations, we find the expressions for matter density and pressures in modified gravity as,
\begin{eqnarray}
\rho &=& \frac{3}{4}\Big[\frac{4B_g}{3} + \frac{a + B + (2 b + a B) r^2 + b B r^4}{(\gamma + 4 \pi) {\Psi}^2}\Big], \label{l1}\\
p_r&=& -{B_g} + \frac{a + B + (2 b + a B) r^2 + b B r^4}{4 (\gamma + 4 \pi) {\Psi}^2},\label{l2}\\
p_t &=& \frac{1}{4 (\gamma + 4 \pi) (3 \gamma + 4 \pi)} \bigg[ 4 B_g (\gamma + 4 \pi)^2 \nonumber\\&&
 + \frac{1}{\Psi}\Big[-B (\gamma - 12 \pi) - 2 a (\gamma + 4 \pi) \nonumber\\&& + 2 (-b + B^2) (\gamma + 4 \pi) r^2\Big] +\frac{1}{{\Psi}^2}\Big\{4 B (\gamma + 4 \pi) \nonumber\\&& - 2 b (7 \gamma + 12 \pi) r^2 +
 a \Big( 2 B (\gamma + 4 \pi) r^2 \nonumber\\&& -(7 \gamma + 12 \pi)\Big)\Big\}  \bigg].\label{l3}
\end{eqnarray}
The anisotropic factor $\Delta=p_t-p_r$ in modified gravity is obtained as,
\begin{eqnarray}\label{del5}
  \Delta &=& \frac{1}{2 (3 \gamma + 4 \pi)} \bigg[ 8{ B_g} (\gamma + 2 \pi)+ \frac{1}{(\gamma + 4 \pi) {\Psi}^2}\times\nonumber\\&&\Big\{(2 B-5a) \gamma +(B-a)8\pi+ \Big(a B (\gamma + 4 \pi) \nonumber\\&&- 2 b (5 \gamma + 8 \pi)\Big) r^2\Big\}+\frac{1}{(\gamma + 4 \pi)\Psi}\Big\{2 B (2 \pi-\gamma) \nonumber\\&&- a (\gamma + 4 \pi)- (b - B^2) (\gamma + 4 \pi) r^2 \Big\}\bigg].
\end{eqnarray}
So, the model parameters like density, radial and transverse pressure, anisotropic factor, electric field etc. in the background of General relativity as well as in modified gravity have been successfully obtained. \par
To find different constants in TK metric potentials, we match our interior solution to the exterior spacetime smoothly outside the event horizon $r>M+\sqrt{M^{2}-Q^2}$, where, $M$ and $Q$ are respectively the total mass and charge enclosed within the boundary $r=R$. Now one can note that the exterior spacetime is zero as there is no matter in the vacuum spacetime. Therefore, there will be no change of GR solution
for the exterior spacetime metric even in the $f(R,\,T )$ theory of gravity. The exterior space-time of the star will be described by the Reissner-Nordstr\"om metric \cite{rn1,rn2} given by
\begin{eqnarray}
ds^{2} &=& -\left(1 - \frac{2M}{r} + \frac {Q^2}{r^2}\right)dt^2 + \left(1 - \frac{2M}{r} + \frac {Q^2}{r^2}\right)^{-1}dr^2
\nonumber\\
&& + r^2(d\theta^2+\sin^2\theta d\phi^2), \label{eq22}
\end{eqnarray}
Now the continuity of the metric coefficients $g_{tt}$, $g_{rr}$ and $\frac{\partial g_{tt}}{\partial r}$ across the boundary surface $r= R$ between the interior and the exterior regions give the following set of relations:
\begin{eqnarray}
1 - 2\tilde{x}+ \tilde{y} &=& C^2 e^{BR^2},\label{eq23}\\
1 - 2\tilde{x} + \tilde{y} &=& (1 + a R^2 + b R^4)^{-1},\label{eq24}\\
\tilde{x} - \tilde{y} &=& B R^2 C^2 e^{BR^2}.\label{eq25}
\end{eqnarray}
where, $\tilde{x}=\frac{M}{R}$ and $\tilde{y}=\frac {Q^2}{R^2} $ and both $\tilde{x},\,\tilde{y}$ are dimensionless quantity.\\
Solving the Eqs.~(\ref{eq23})-(\ref{eq25}) and using the  condition $p_r(R)=0$ , one can determine the values of the constants $B$, $C$, $a$ and $b$ as,
\begin{eqnarray}
B &=& \frac{U}{R^2} \left[\tilde{x} - \tilde{y}\right],\label{eq26}\\
C &=&  e^{-\frac{BR^2}{2}} U^{-\frac{1}{2}}, \label{eq27}\\
a &=& \frac{1}{(\gamma + 2 \pi) R^2 (1 + 3U^2)}\Big[-2 (\gamma + 2 \pi) + (\gamma + 2 \pi) \times\nonumber\\&&(2 +  B R^2)U+6 B \pi U^2 R^2 \Big], \label{eq28}\\
b &=& \frac{-1 - aR^2 + U}{R^4}
\end{eqnarray}
where, $ U = \Big(1 - 2\tilde{x} + \tilde{y}\Big)^{-1}$.\\
Now from the condition that the radial pressure vanishes at the boundary of the star $(p_r(r=R)=0)$ one can get,
\begin{equation}\label{bn2}
\rho_s = 4 B_g,
\end{equation}
where $\rho_s$ is the surface density given by,
\begin{eqnarray}
\rho_s&=&  B_g + \frac{3 \big(a + B + (2 b + a B) R^2 + b B R^4\big)}{4 (\gamma + 4 \pi) (1 + a R^2 + b R^4)^2},
\end{eqnarray}
For drawing the plots we have considered the compact star SAX J $1808.4-3658$ by assuming $M = 0.88~M_{\odot},\, R = 8.9 $ km. Along with this we have also assumed $Q = 0.0089$. The central density, surface density and central pressure for different values of $\gamma$ have been obtained in table~\ref{tbl2}.

\section{Physical aspects of $f(R,\,T)$ gravity}\label{sec7}
In this section we perform both analytic and graphical analysis
in order to check the physical and mathematical properties
of our present model. Now we shall check the conditions one by one.
\begin{itemize}
\item {\bf Metric potentials :} In this paper, we choose the metric potentials as,
$e^{\nu}=C^2 e^{Br^2},~~e^{\lambda}=(1 + a r^2 + b r^4)$,
we note that $e^{\nu}|_{r=0}=e^{C}>0$ and $e^{\lambda}|_{r=0}=1$, moreover,
$\left(e^{\nu}\right)'=2 B e^{C + B r^2} r ,~~\left(e^{\lambda}\right)'=2 A e^{A r^2} r$. We
will notice in the later section that this behavior allows to match the inner geometry to the exterior space-time in a smoothly way at the
boundary $r=R$ to get the constant parameters that characterize the model. So the metric potentials are well behaved in the interval (0,\,R).

\item {\bf Pressure and density :} In this regard the main thermodynamic variables must
respect some criteria. From expressions (\ref{l1})-(\ref{l3}), at the center of the compact configuration we have,
\begin{eqnarray*}
\rho_c&=& \frac{3 a + 3 B + 4 { B_g} (\gamma + 4 \pi)}{4 (\gamma + 4 \pi)},\\
 p_c&=& -{ B_g}+\frac{a + B}{4 (\gamma + 4 \pi)},
\end{eqnarray*}
where $\rho_c$ and $p_c$ are respectively the central density and central pressure of the compact star in modified gravity. The corresponding quantities in GR can be obtained by simply putting $\gamma=0$.\\
We obtain the density and pressure gradients for our present model by taking the differentiation of the eqns.~(\ref{l1})-(\ref{l3}) with respect to r as,
\begin{eqnarray*}
\rho'&=& -\frac{3r}{2 (\gamma + 4 \pi){\Psi}^3}\Big[a^2 (2 + B r^2) + a (B \nonumber\\ && + 6 b r^2 + 3 b B r^4) + 2 b \big\{-1 + 3 b r^4 \nonumber\\ &&+ B (r^2 + b r^6)\big\}\Big]<0,\label{k1}\\
      p_r'&=& -\frac{r}{2 (\gamma + 4 \pi){\Psi}^3}\Big[a^2 (2 + B r^2) + a (B \nonumber\\ && + 6 b r^2 + 3 b B r^4) + 2 b \big\{-1 + 3 b r^4 \nonumber\\ &&+ B (r^2 + b r^6)\big\} \Big]<0,\label{k2}\\
      p_t'&=& \frac{r}{2 (\gamma + 4 \pi) (3 \gamma + 4 \pi) {\Psi}^3} \Big[2\Psi \big\{a B (\gamma + 4 \pi) \nonumber\\ &&- b (7 \gamma + 12 \pi)\big\} -\Psi (a + 2 b r^2) \big\{B (12 \pi-\gamma) \nonumber\\ &&- 2 a (\gamma + 4 \pi) +2 (-b + B^2) (\gamma + 4 \pi) r^2\big\}  \nonumber\\ && +2{\Psi}^2 (-b + B^2) (\gamma + 4 \pi)-2 (a + 2 b r^2)\times
        \nonumber\\ &&  \Big\{4 B (\gamma + 4 \pi) - 2 b (7 \gamma + 12 \pi) r^2 \nonumber\\ &&-
  a \big\{7 \gamma + 12 \pi -2 B (\gamma + 4 \pi) r^2\big\}\Big\} \Big]<0.  \label{k3}
\end{eqnarray*}
We see that \[\rho'|_{r=0}=0,\,p_r'|_{r=0}=0,\,p_t'|_{r=0}=0.\]
and,
\begin{eqnarray*}
\rho''|_{r=0}&=& \frac{6 b - 3 a (2 a + B)}{2 (\gamma + 4 \pi)},\\
p_r''|_{r=0}&=& \frac{2 b - a (2 a + B)}{2 (\gamma + 4 \pi)},\\
p_t''|_{r=0}&=& \frac{-2 b + a (2 a + B)}{(\gamma + 4 \pi)}\\&& + \frac{4 a^2 - 4 b - 11 a B + 2 B^2}{ (6 \gamma + 8 \pi)}.
\end{eqnarray*}
\begin{figure}[htbp]
    \centering
        \includegraphics[scale=.45]{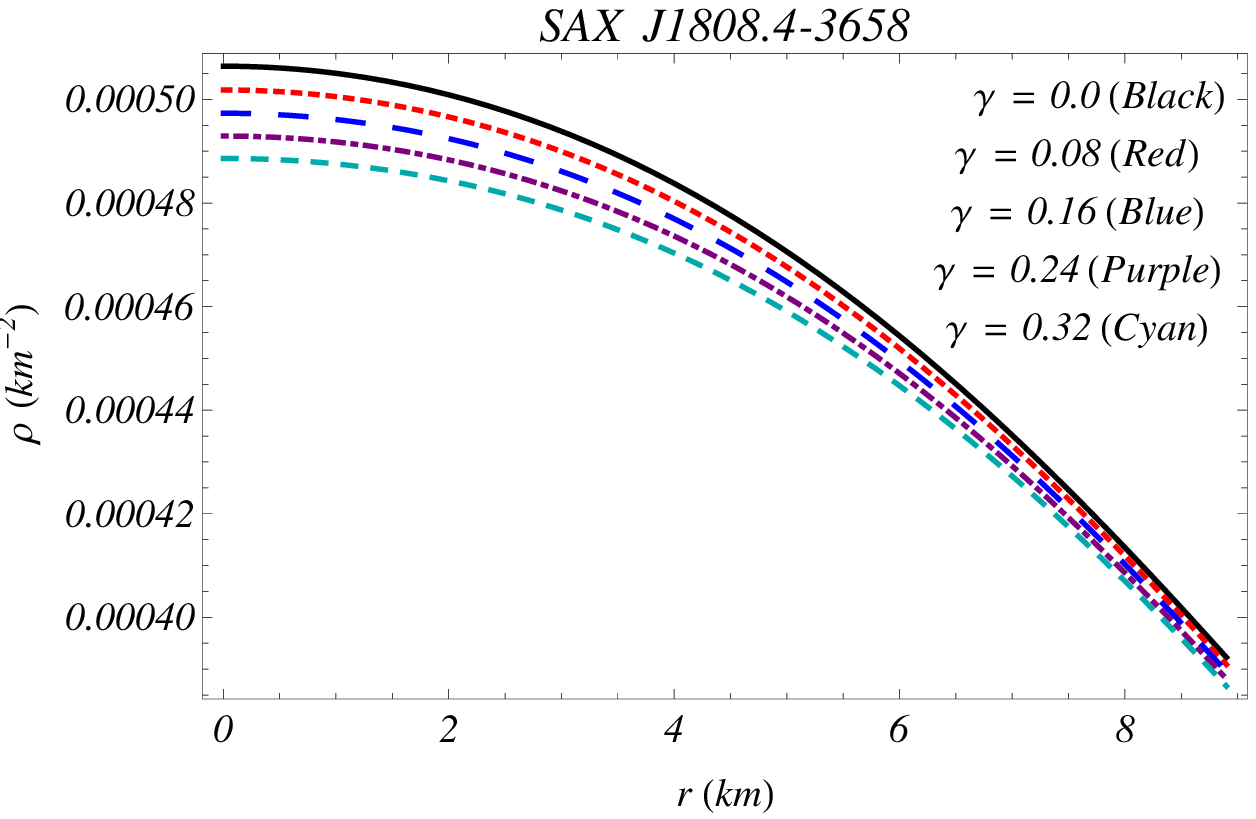}
        \includegraphics[scale=.45]{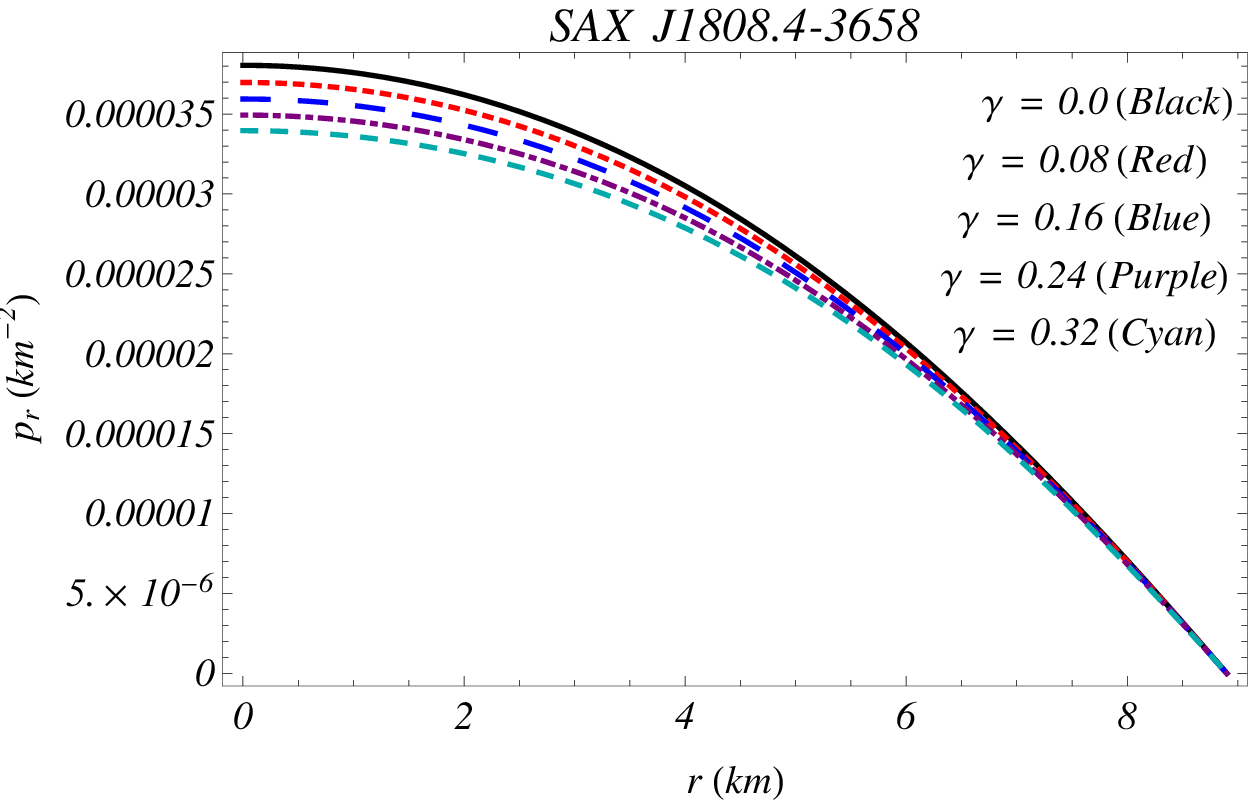}
        \includegraphics[scale=.45]{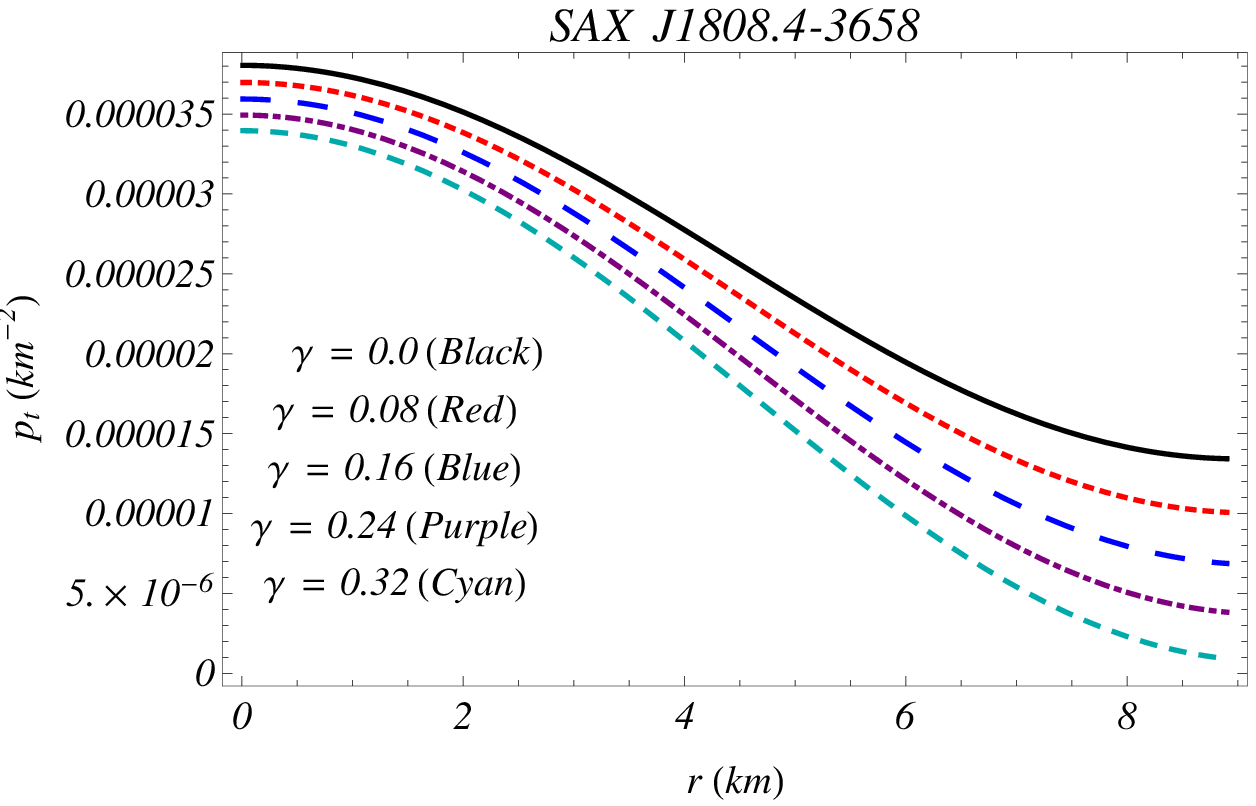}
       \caption{(Top) Matter density $\rho$, (middle) radial pressure $p_r$, (bottom) transverse pressure $p_t$ are plotted against $r$ inside the stellar interior for different values of $\gamma$ mentioned in the figure. \label{pr8}}
\end{figure}
The profiles of the matter density and both the pressures are plotted against radius in Fig.~\ref{pr8}. The profiles show that $\rho,\,p_r$ and $p_t$ all are positive for $r \in (0,\,R)$. All are monotonic decreasing functions of $r$, i.e., all of them take maximum value at the center of the star and take minimum value at the boundary. What is more that the radial pressure vanishes at the boundary of the star but both transverse pressure and matter density are positive at the boundary of the star.\par
For a physically acceptable model, the pressure should be non negative inside the fluid sphere, and therefore, $p_c>0~ \Rightarrow~ \frac{a+B}{4(\gamma+4\pi)}>B_g$. Again by Zeldivich \cite{zel} condition $p_c/\rho_c~<1~\Rightarrow~B_g>-\frac{a+B}{4(\gamma+4\pi)}$. This lower limit for bag constant $B_g$ is always satisfied since $B_g$ is positive. So we obtain a reasonable upper bound for the bag constant as,
\begin{eqnarray}
B_g<\frac{a+B}{4(\gamma+4\pi)}.
\end{eqnarray}
\item {\bf Causality Condition :} In order to fulfill the physical requirements for realistic models, it is necessary to examine the causality and hydrostatic equilibrium of the present self-gravitating system. First, we
discuss the causality condition of the model which says that
the velocity of sound must be less than the velocity of light
everywhere within the object. The square of radial and transverse velocity of sound $V_r^2$ and $V_t^2$ respectively are given by,
\begin{eqnarray}
V_r^2= \frac{dp_r}{d\rho}&=&\frac{1}{3},\\
V_t^2=\frac{dp_t}{d\rho}&=& \frac{1}{\chi}\Big[-16 a^2 \gamma + 16 b \gamma + 5 a B \gamma - 2 B^2 \gamma -
  32 a^2 \pi \nonumber\\
  &&+ 32 b \pi + 36 a B \pi - 8 B^2 \pi + D_1r^2
 -3 b D_2 r^4
  \nonumber\\
  && +2 b D_3 r^6 - 2 b^2 (b - B^2) (\gamma + 4 \pi) r^8\Big],
\end{eqnarray}
where $\chi=\chi(r)$ and its expression is given by,
\begin{eqnarray*}
\chi&=&3 (3 \gamma + 4 \pi) \Big[a^2 (2 + B r^2)
  + a (B + 6 b r^2 + 3 b B r^4) \\&&+  2 b \big(-1 + 3 b r^4
   + B (r^2 + b r^6)\big)\Big],
  \end{eqnarray*}
and $D_1,\,D_2$ and $D_3$ are constants and depend on $\gamma$ and their expressions are as follows:
  \begin{eqnarray*}
  D_1&=&-2 a^3 (\gamma + 4 \pi) + a^2 B (\gamma + 20 \pi) +2 b B (7 \gamma + 44 \pi) \nonumber\\&&- 2 a \big(B^2 (\gamma + 4 \pi) + b (22 \gamma + 40 \pi)\big), \\
  D_2&=&2 a^2 (\gamma + 4 \pi) + 2 b (7 \gamma + 12 \pi) - a B (\gamma + 20 \pi),\\
  D_3&=&-b B (\gamma - 12 \pi) - a (3 b - B^2) (\gamma + 4 \pi)
  \end{eqnarray*}

\begin{figure}[htbp]
    \centering
        \includegraphics[scale=.45]{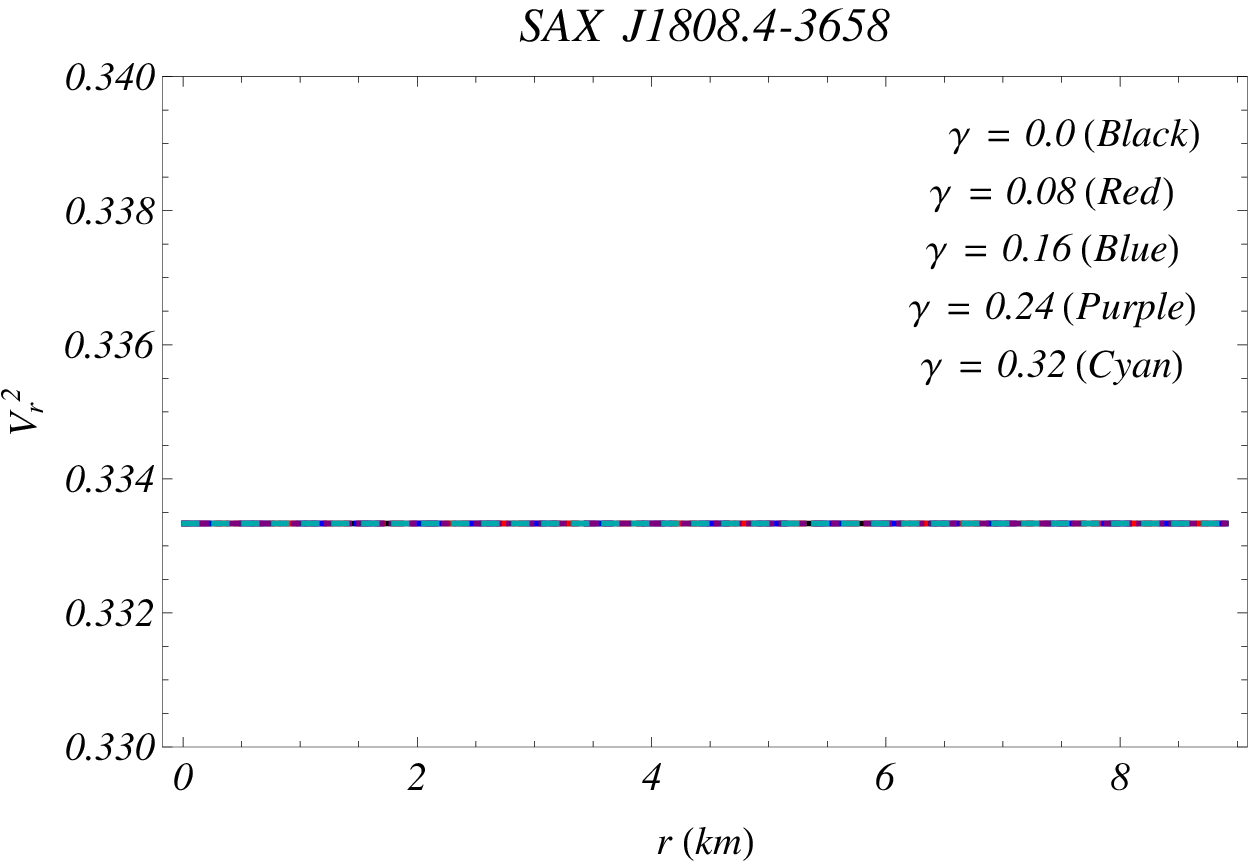}
        \includegraphics[scale=.45]{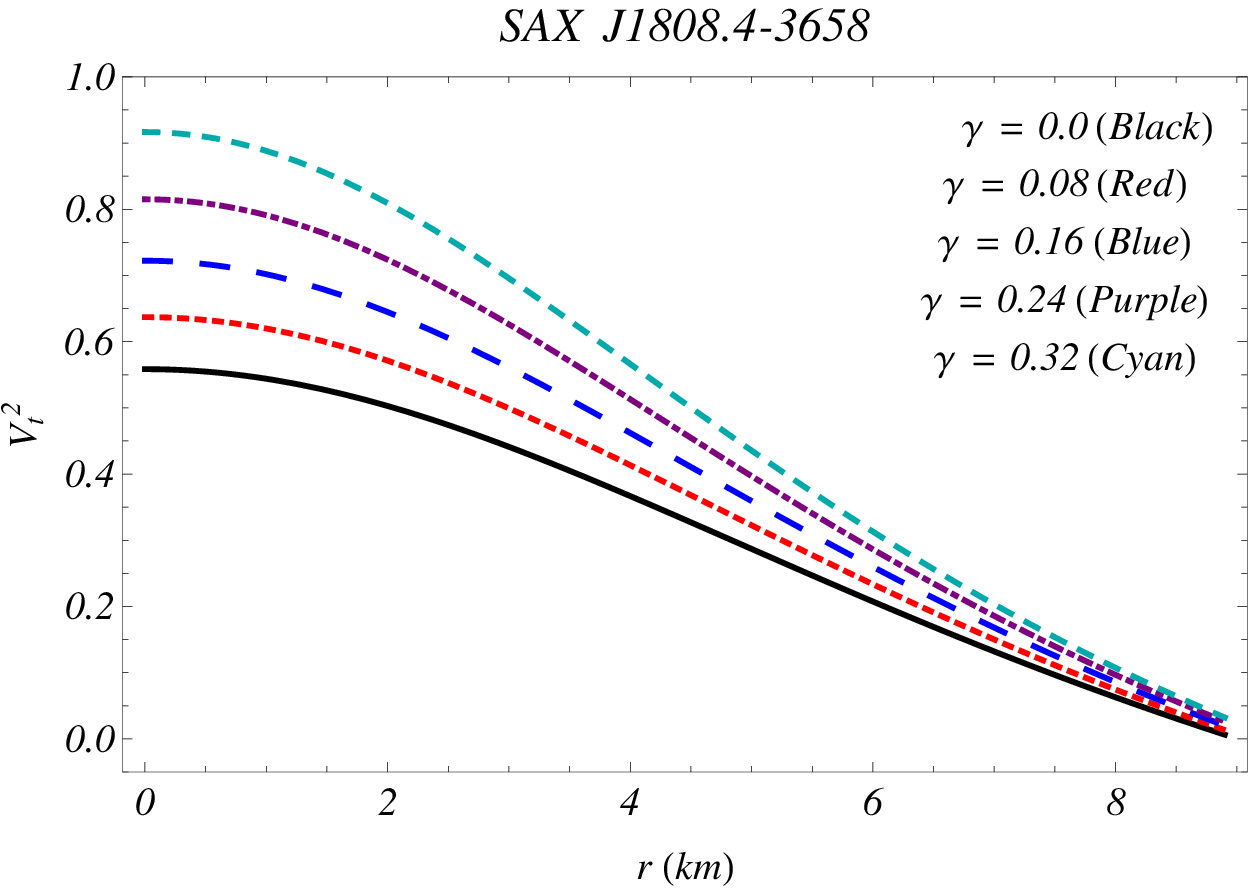}
        \includegraphics[scale=.45]{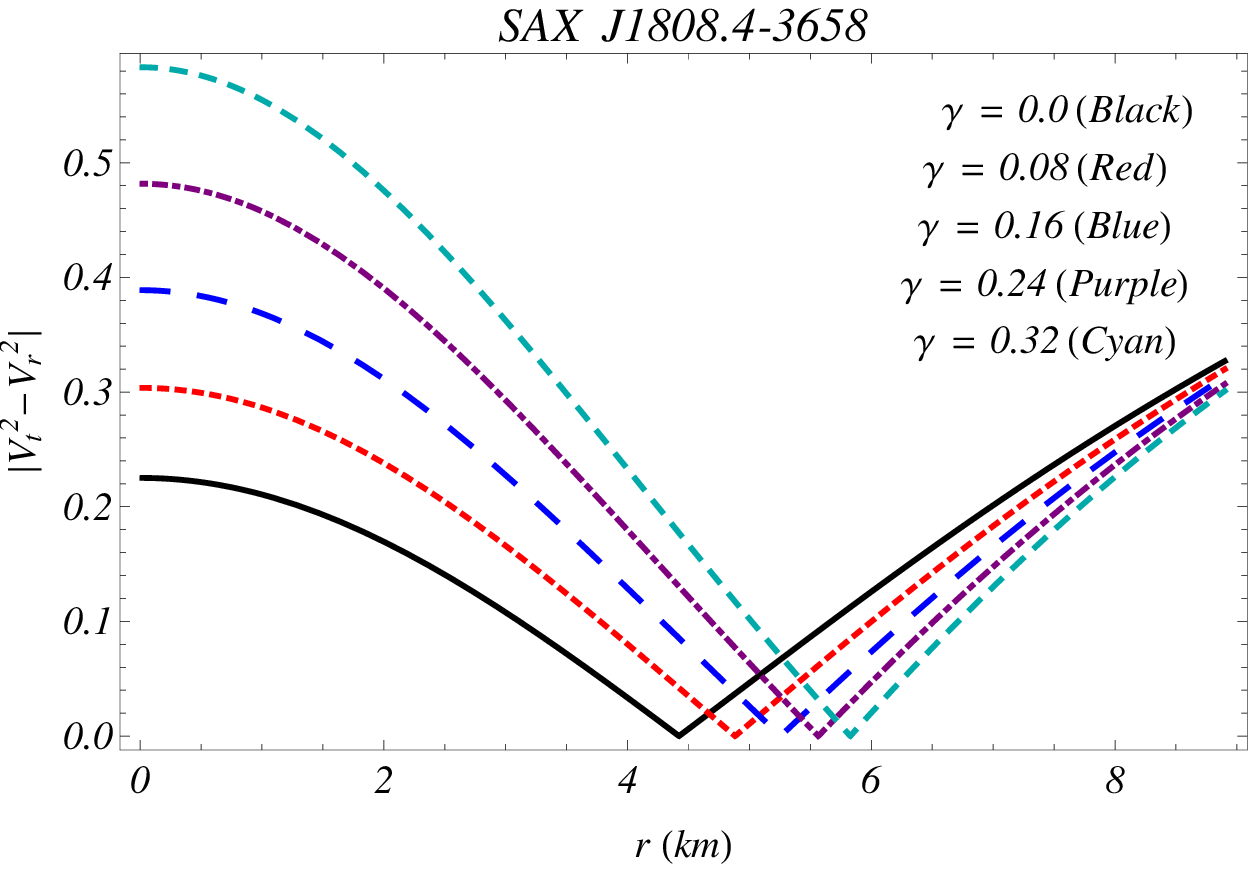}
       \caption{(Top) Square of the radial sound velocity $V_r^2$, (middle) square of the transverse sound velocity $V_t^2$ and (bottom) the stability factor $|V_t^2-V_r^2|$ is plotted against r for the strange star candidate SAX J1808.4-3658 by taking different values of $\gamma$.\label{sv}}
\end{figure}
Moreover for a relativistic object, Herrera \cite{her47} proposed the method of ``cracking'' which is related to the
stability of anisotropic stars under small radial perturbations. Using the concept of cracking, Abreu et al. \cite{abreau}
proved that the region of an anisotropic star where the radial speed of sound crosses the transverse speed of sound is potentially stable, otherwise, the region is potentially unstable. In mathematical terms, it can be written as,
\begin{eqnarray*}
-1 \leq V_t^2-V_r^2\leq 0~\Rightarrow~\text{Potentially stable region}\\
0 < V_t^2-V_r^2 \leq 1~\Rightarrow~\text{Potentially unstable region}
\end{eqnarray*}
Since both sound speeds maintain causality, we have, $V_r^2,\,V_t^2<1$ \cite{ha1,ha2}. Again, Le Chatelier's principle requires that $V_r^2,\,V_t^2>0$. Combining the above two cases we get $0<V_r^2,\,V_t^2<1$ and it gives, $-1<V_r^2-V_t^2<1$, it further implies, $|V_r^2-V_t^2|<1$. The profiles of $V_r^2,\,V_t^2$ and $|V_r^2-V_t^2|$ for different values of $\gamma$ are plotted in Fig.~\ref{sv}.\par
  \item {\bf Relativistic Adiabatic index :} To study the stability of both relativistic and non-relativistic compact stars model, relativistic adiabatic index is used and it also characterizes stiffness of the EoS for a given density. After the pioneering
work by Chandrasekhar \cite{chandras}, later on many scientists studied dynamical stability
of the stellar system against an infinitesimal radial adiabatic perturbation. The collapsing condition of anisotropic model is given by $\Gamma_r<4/3$ \cite{bondi}. The radial adiabatic index $\Gamma_r$ reads as \cite{chan}
\begin{eqnarray}
\Gamma_r=\left(\frac{\rho}{p_r}+1\right)\left(\frac{dp_r}{d\rho}\right)_S,
\end{eqnarray}
\begin{figure}[htbp]
    \centering
        \includegraphics[scale=.45]{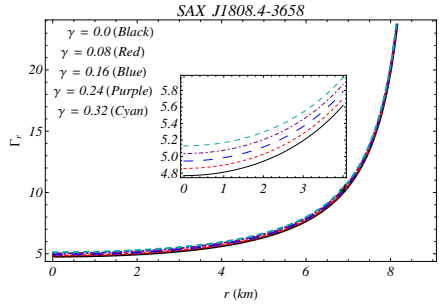}
       \caption{The relativistic adiabatic index $\Gamma_r$ has been plotted against r inside the stellar interior \label{gam1}.}
\end{figure}
where the derivation is performed at constant entropy $S$ and $\frac{dp_r}{d\rho}$ is the speed of sound in units of the speed
of light. From the expression of $\Gamma_r$ we see that adiabatic index depends on that ratio $\rho/p_r$. The profile of $\Gamma_r$ has been plotted in Fig~\ref{gam1} for different values of $\gamma$.

\item {\bf Behavior of electric field and anisotropic factor :} The expression of the electric field is given in eqn. (\ref{k4}). Now for a physically acceptable model it is required that $E^2(r=0)=0$ which leads the following equation :
    \begin{eqnarray}
    \frac{3 a \gamma - 4 \gamma^2 B_g + 6 a \pi - 6 B \pi - 24 \gamma B_g \pi -
 32 B_g \pi^2}{3 \gamma + 4 \pi}=0 \nonumber\\
    \end{eqnarray}
    The above eqn. gives the expression of the bag constant as,
    \[B_g=\frac{3}{4}\left(-\frac{B}{\gamma+2\pi} + \frac{a+B}{\gamma+4\pi}\right).\]
    The expression of the anisotropic factor is given in eqn.~(\ref{del5}) which is defined as the difference between the transverse and radial pressure. It may be positive or negative according as $p_t>p_r$ or $p_t<p_r$ and it is denoted by $\Delta$. The anisotropic force is defined by $\frac{2\Delta}{r}$, and this force may be positive or negative depending upon the sign of $\Delta$, this force vanishes for isotropic case. The profiles of electric field and anisotropic factor is plotted in Fig.~\ref{ec3}
    \begin{figure}[htbp]
    \centering
        \includegraphics[scale=.45]{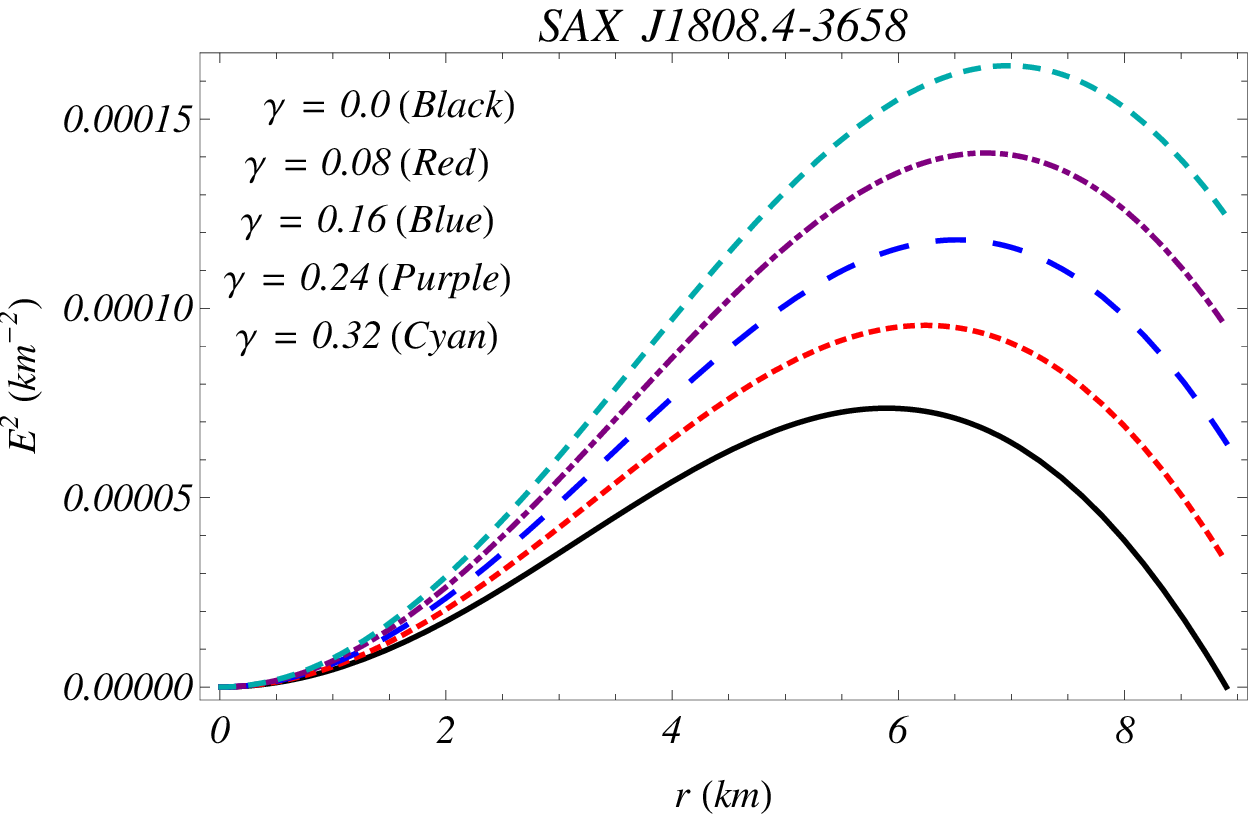}
        \includegraphics[scale=.45]{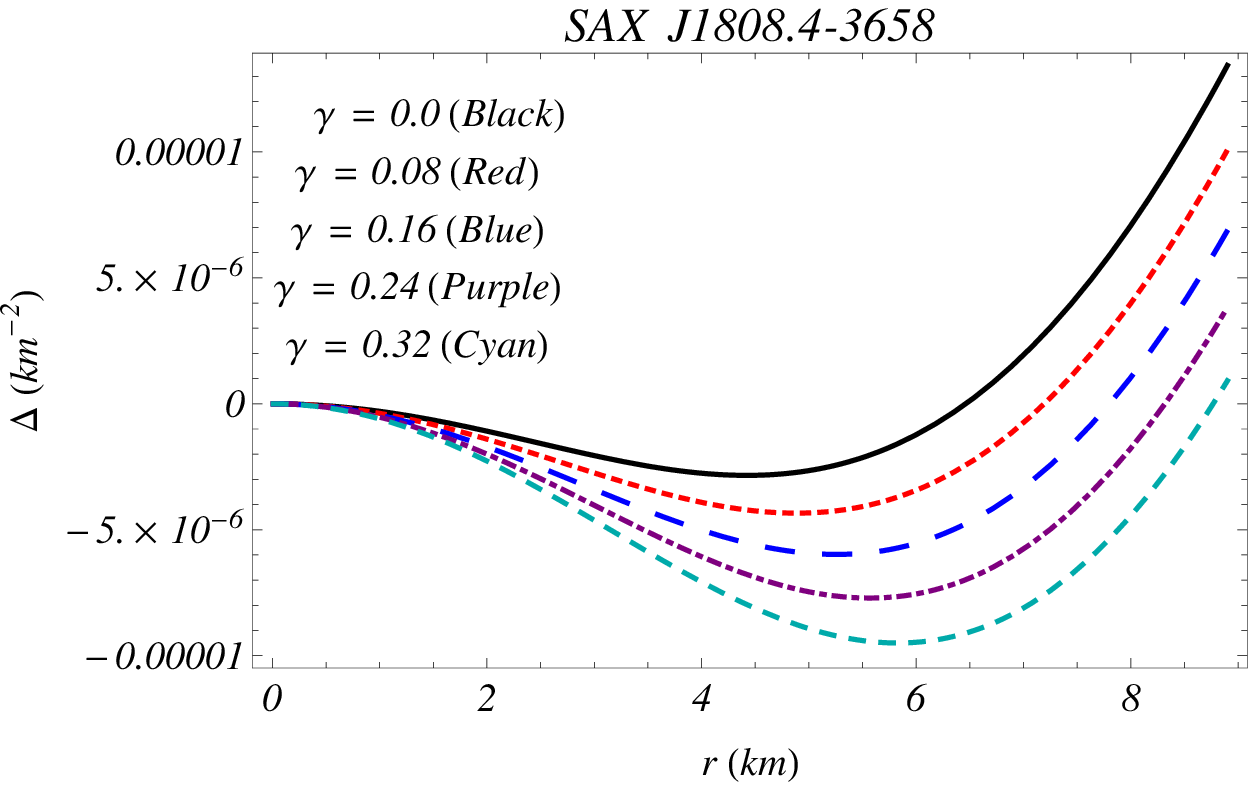}
       \caption{(Top) $E^2$ and (bottom) anisotropic factor $\Delta$ are shown against radius for different values of $\gamma$ mentioned in the figure.\label{ec3}}
\end{figure}

\item {\bf Energy conditions :} Our proposed model of charged compact compact star will satisfy Null energy condition (NEC), Weak Energy Condition (WEC), Strong Energy Condition (SEC) and Dominant energy condition (DEC) if the following inequality hold simultaneously for each and every point inside the stellar model:
\begin{itemize}
\item NEC:~$\rho+p_r \geq 0,~\rho + p_t +\frac{E^2}{4\pi} \geq 0,$ and this energy condition implies that
an observer crossing a null diagram will quantify the usual matter density to be nonnegative.
\item WEC:~$\rho+p_r \geq 0,~\rho + p_t +\frac{E^2}{4\pi} \geq 0,~ \rho + \frac{E^2}{8\pi} \geq 0,$ and WEC suggests that the matter density
measured by an observer traversing a time-like diagram is always positive.
\item SEC:~$\rho+p_r \geq 0,~\rho + p_t +\frac{E^2}{4\pi} \geq 0, \rho+ p_r +2 p_t+ \frac{E^2}{4\pi} \geq 0,$ SEC indicates that the trace of the tidal tensor tested by the relating observers is always positive.
\item DEC:~$\rho-p_r +\frac{E^2}{4 \pi}\geq 0,~\rho - p_t \geq 0,~ \rho + \frac{E^2}{8\pi} \geq 0$, DEC represents mass-energy that can never be seen to be flowing faster than light.
\end{itemize}

The expressions on the l.h.s of the above inequalities have been plotted in Fig.~\ref{ec9}. From the figure we see that all the energy conditions are satisfied. In our study on the charged compact star, SEC is satisfied implying
the fact that gravity will be attractive and also the matter-energy density will be
always positive.

\item {\bf Equation of state :} The equation of state parameters $\omega_r$ and $\omega_t$ can be obtained from the following relation: \[p_r=\omega_r \times \rho,~~p_t=\omega_t \times \rho.\]
      Moreover, we know that the radial pressure maintains a linear relationship from our assumptions but the variation of transverse pressure with respect to density is unknown to us. The Equation of state parameter and variation of pressure with respect to density is depicted in Fig.~\ref{eos}.

      \item{\bf Mass, compactness and Redshift :}
The effective gravitational mass within the radius `r' of the charged strange star can be obtained from the following formula \cite{murad}
\begin{eqnarray}\label{m68}
m(r)&=&4\pi \int_0^r \rho(\tilde{r})~\tilde{r}^2 d\tilde{r} +\frac{q^2}{2r}+\frac{1}{2}\int_0^r\frac{q(\tilde{r})^2}{\tilde{r}^2}d\tilde{r},\nonumber\\
&=& m^{\text{eff}} -\frac{\gamma}{2}\int_0^r(\rho-p_r-2p_t)(\tilde{r})~\tilde{r}^2 d\tilde{r}\nonumber\\&& +\frac{q^2}{2r}+\frac{1}{2}\int_0^r\frac{q(\tilde{r})^2}{\tilde{r}^2}d\tilde{r},
\end{eqnarray}
In equation (\ref{m68}), $m^{\text{eff}}=4\pi \int_0^r \rho^{\text{eff}} (\tilde{r})\tilde{r}^2 d\tilde{r}$. In Einstein-Maxwell gravity the mass function is obtained when $\gamma~\rightarrow~0$. The mass function inside the radius `r' of the charged fluid sphere can be obtained as,
\begin{eqnarray}
m(r)&=& \frac{r}{2} \Big(1 - \frac{1}{\Psi}\Big) + \frac{r^3}{2 (3 \gamma + 4 \pi)}
    \bigg[-4 ~{ B_g}\times\nonumber\\ && (\gamma + 2 \pi) (\gamma + 4 \pi) + \frac{1}{\Psi}\Big[2 a (\gamma + 2 \pi) - 2 B (\gamma + 3 \pi) \nonumber\\&& +\Big(B^2 \gamma + 2 b (\gamma + 2 \pi)\Big) r^2\Big] + \frac{1}{{\Psi}^2}\Big[2 B \gamma +2 b (\gamma \nonumber\\ && + 2 \pi) r^2
    + a (\gamma + 2 \pi + B \gamma r^2)\Big]\bigg].
\end{eqnarray}
The mass function is regular at the center as $m(r)\rightarrow 0$ as $r\rightarrow 0$.
\begin{figure}[htbp]
    \centering
        \includegraphics[scale=.45]{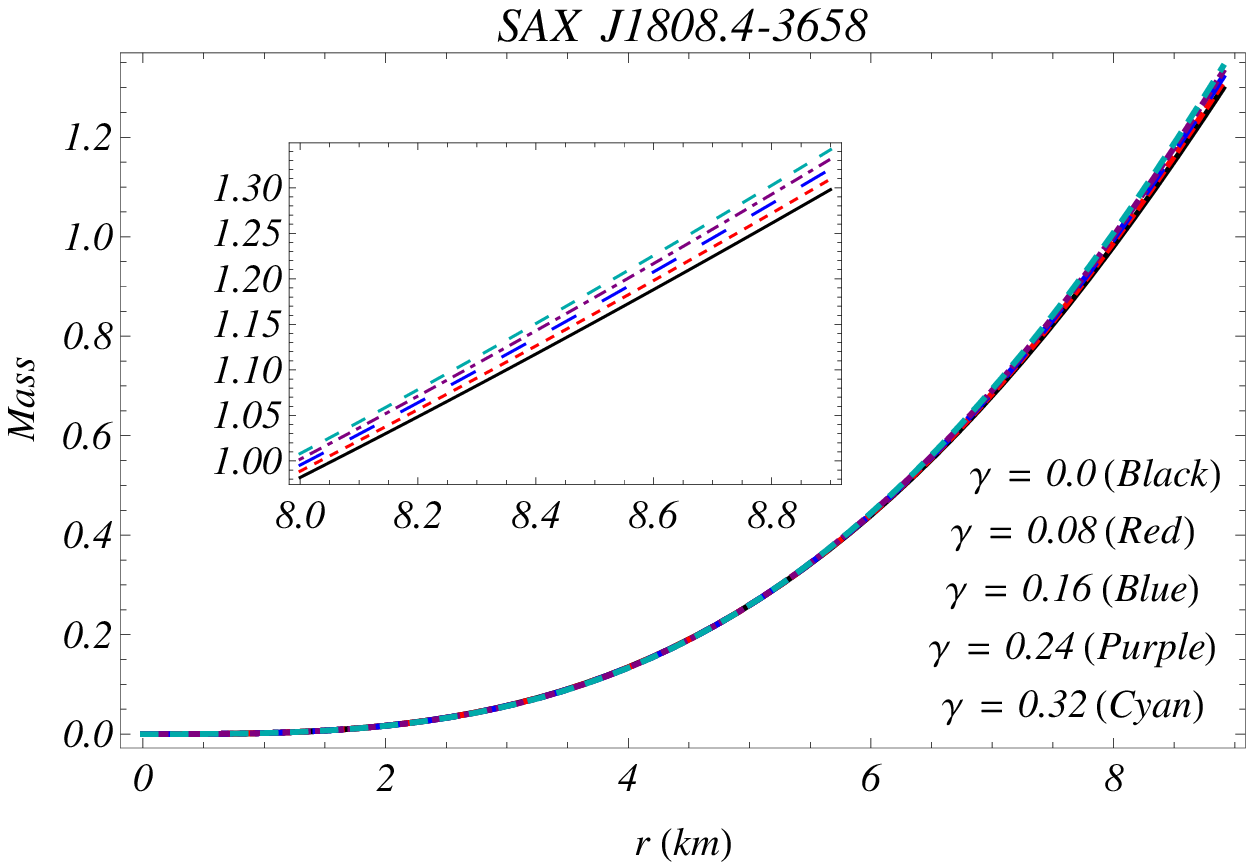}
        \includegraphics[scale=.45]{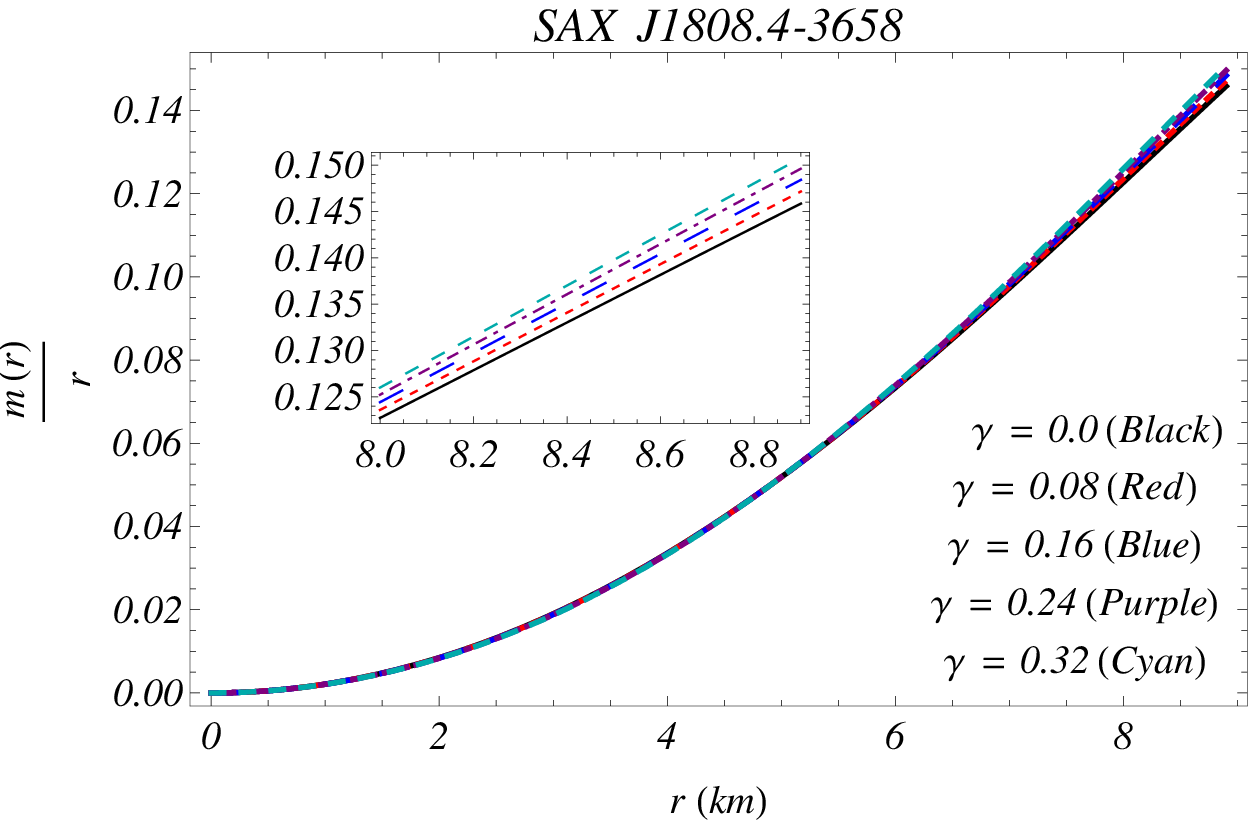}
       \caption{The variation of mass function and compactness are shown against radius for different values of $\gamma$ mentioned in the figure.\label{mass}}
\end{figure}
\begin{figure}[htbp]
    \centering
        \includegraphics[scale=.45]{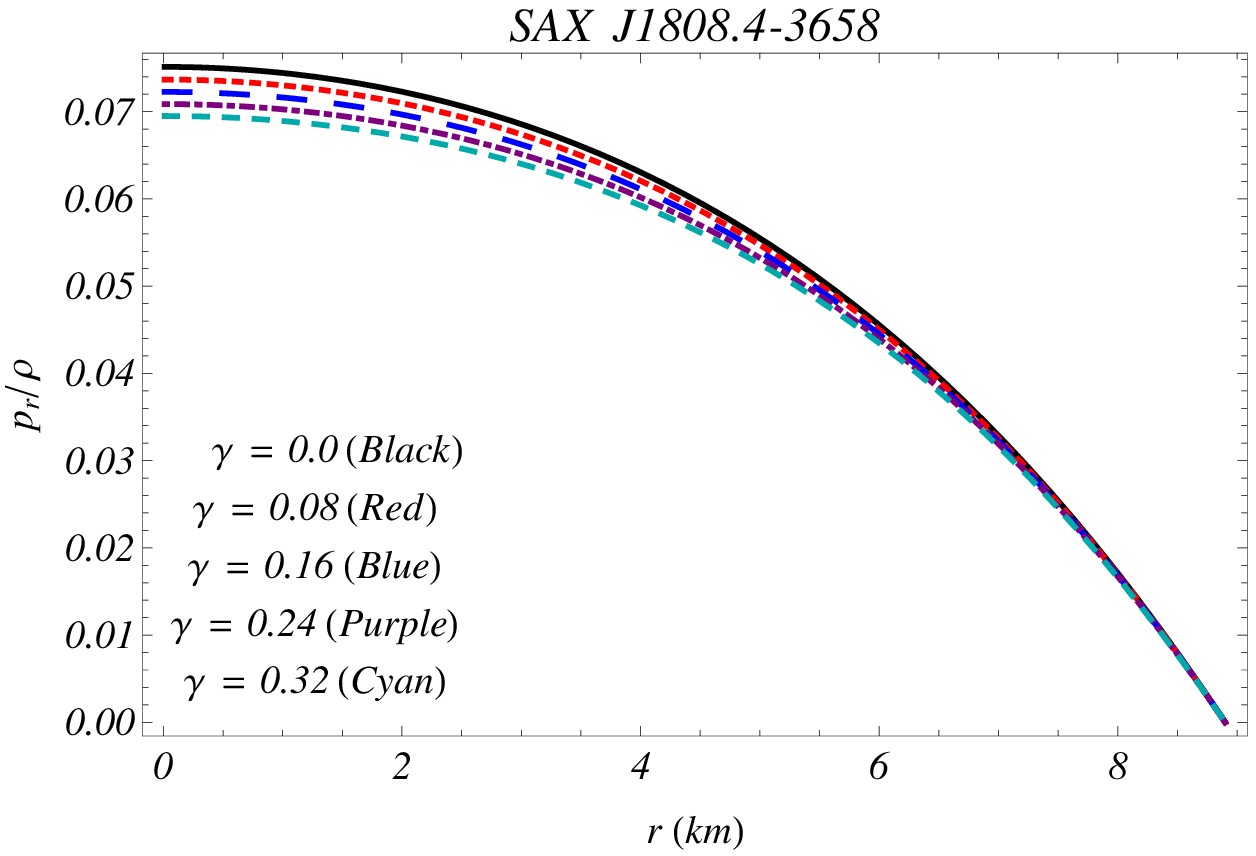}
        \includegraphics[scale=.45]{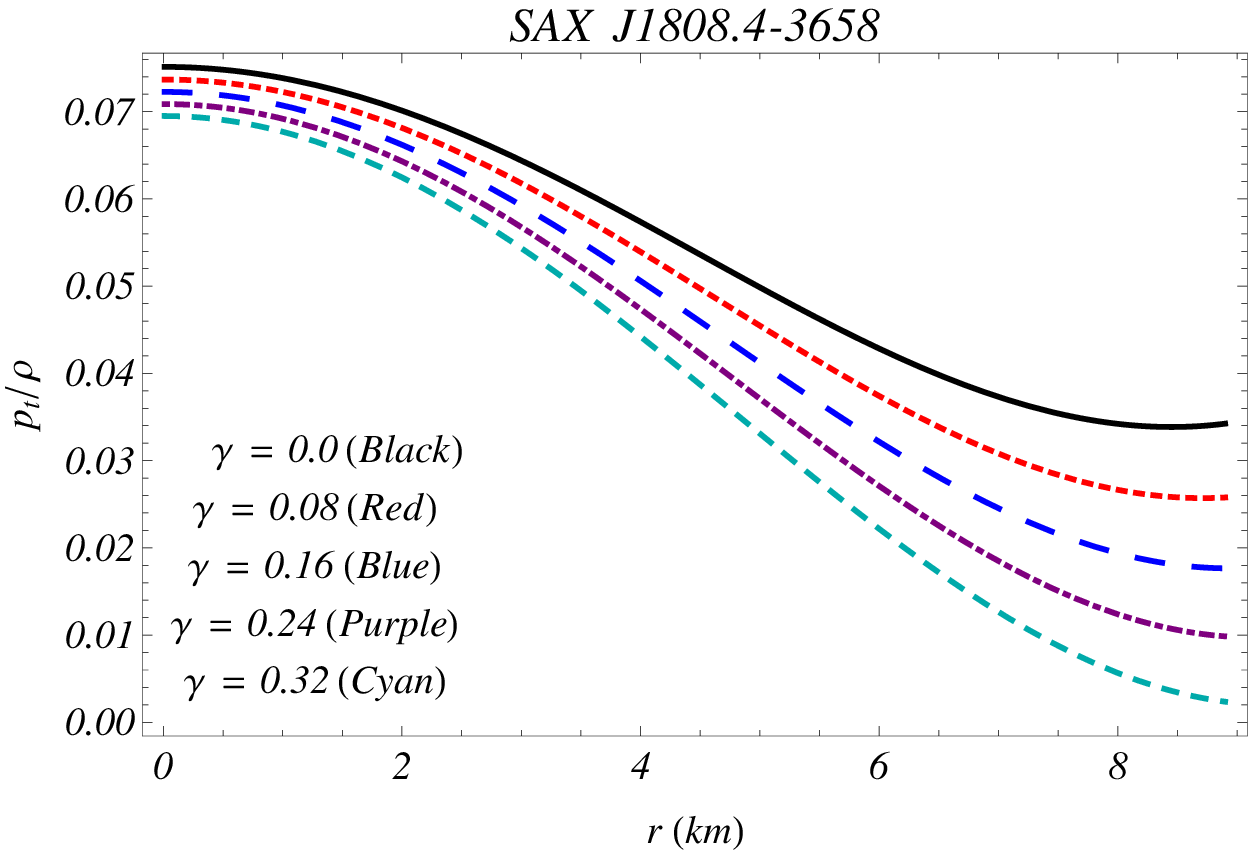}
        \includegraphics[scale=.45]{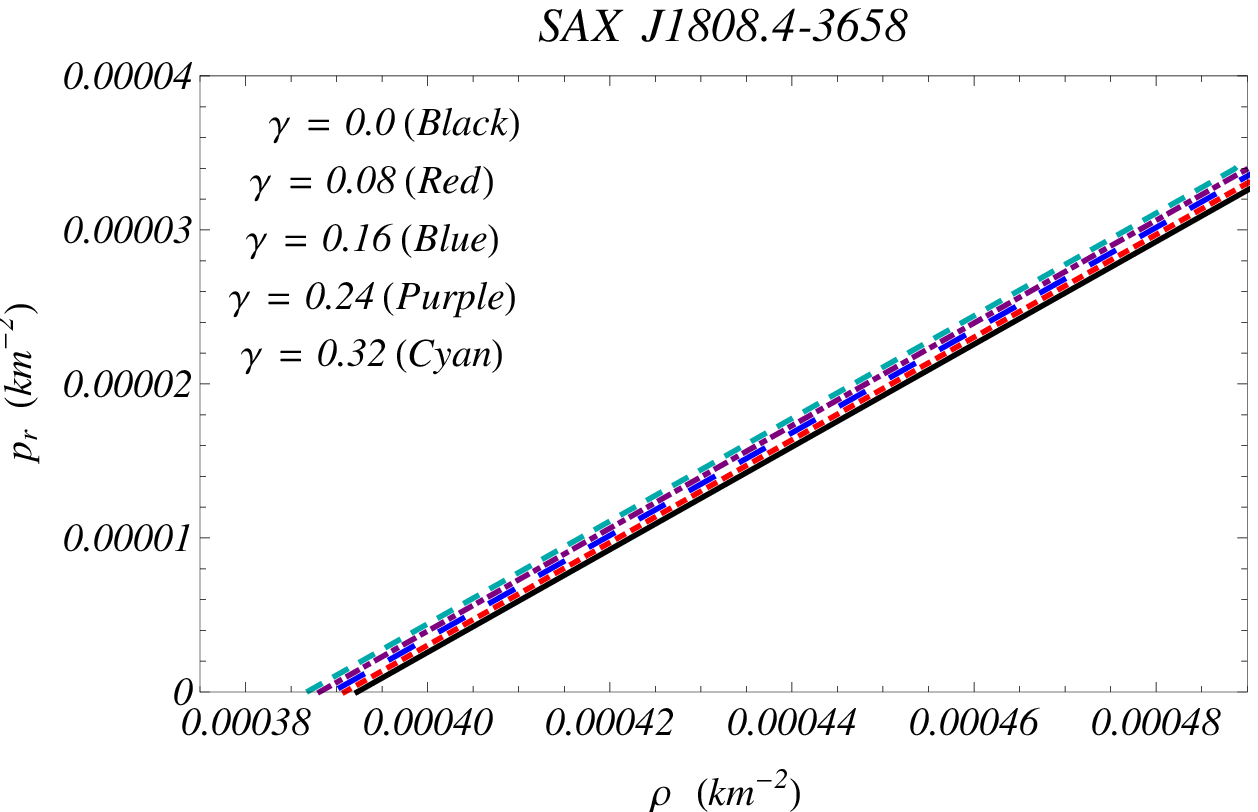}
        \includegraphics[scale=.45]{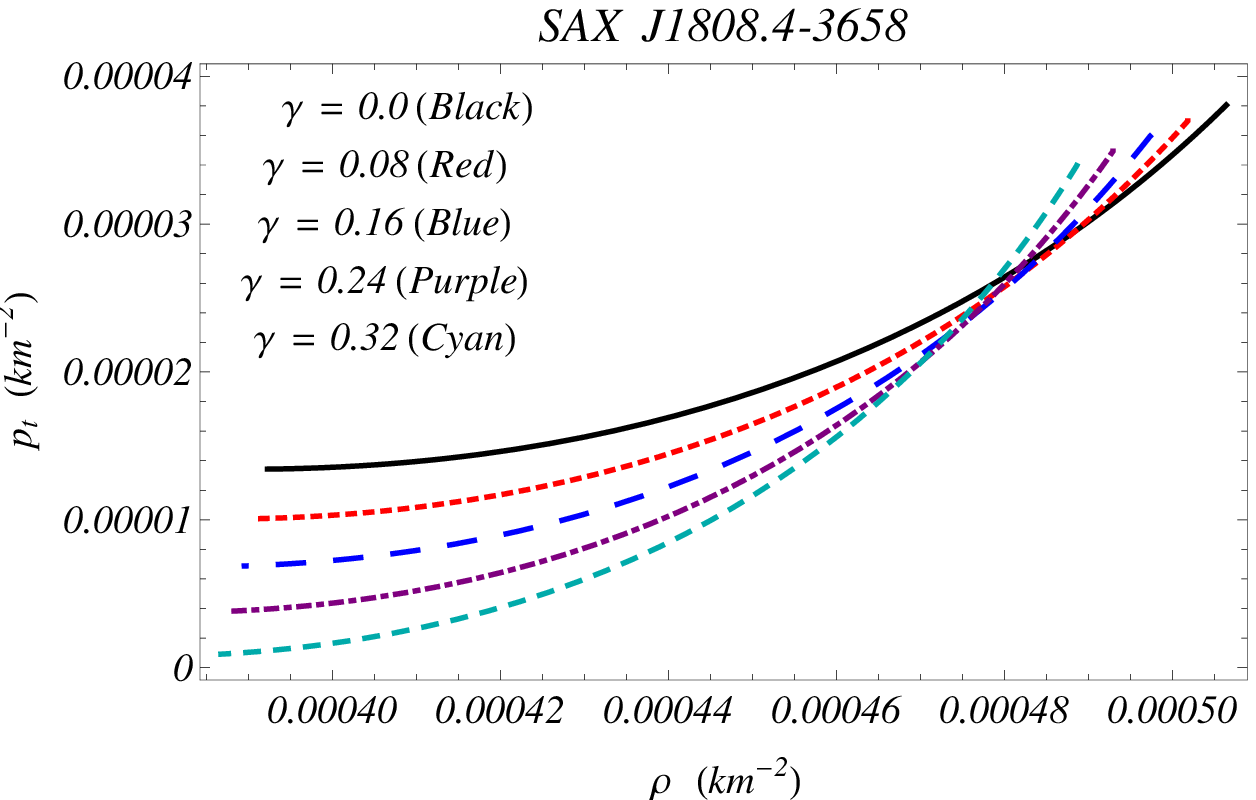}
       \caption{The pressure density relation are shown for different values of $\gamma$ mentioned in the figure.\label{eos}}
\end{figure}
The compactness factor inside the radius `r' for our present model is obtained as,
\begin{eqnarray}
u(r)=\frac{m(r)}{r},
\end{eqnarray}
We denote $\mathcal{U}=\mathcal{M}/R$, where $\mathcal{M}=m(R)$. The compactness factor is useful to classify the compact objects in different category: for normal star $\mathcal{U}~\sim~10^{-5}$, in case of white dwarfs $\frac{M}{R}~\sim~10^{-3}$, for neutron star, $\mathcal{U} \in (10^{-1},\,\frac{1}{4})$, when $\mathcal{U}$ lies between $\left(\frac{1}{4},\,\frac{1}{2}\right)$, it denotes ultra compact star and
if $\mathcal{U}=\frac{1}{2}$, it represents Black hole.\par
For a charged compact star model, B\"{o}hmer
and Harko \cite{harko15} proposed a lower bound for compactness factor, whereas, Andr\'{e}asson \cite{an1} proposed an upper bound for compactness factor. Combining the two results a bound for compactness factor for charged model of compact star is obtained as,
 \begin{eqnarray}\label{7z}
\frac{3Q^2}{4R^2}\frac{1+\frac{Q^2}{18R^2}}{1+\frac{Q^2}{12R^2}} \leq \mathcal{U}\leq \left(\frac{1}{3}+\sqrt{\frac{1}{9}+\frac{Q^2}{3R^2}}\right)^2,
\end{eqnarray}
where, $Q$ is the total charge inside the star, i.e., $q(r=R)=Q$. In case of uncharged compact object $Q=0$ and under this condition the upper limit of $\mathcal{U}$ in Eqn. (\ref{7z}) obeys the Buchdahl's limit \cite{buch} $\mathcal{U}<\frac{4}{9}$. We have shown the nature of mass function and compactness in Fig.~\ref{mass} for different values of $\gamma$ mentioned in the figure.\par
Now we are in a position to check the bound for $\mathcal{U}$ given in Eqn.~(\ref{7z}) for different values of $\gamma$. For this purpose we have find the numerical value of $\mathcal{U}$ from our model which have been presented in the following table and it confirms that the inequality is verified for our model of charged compact star in $f(R,\,T)$ gravity.

\begin{center}
\begin{tabular}{ c | c |c  |c }
$\gamma$ &  Value of lower & $u^{\text{eff}}(R)$&  Value of upper\\
&limit of eq.~(\ref{7z})& &limit of eq.~(\ref{7z})\\
\hline
$0.0$ &$0.0007499$&$0.145843$&$0.666667$\\
$0.08$ &$ 0.0382305$&$0.147145$&$0.667967$\\
$0.16$ &$0.0535169$&$0.148398$&$0.669213$ \\
$0.24$ &$0.0649036$&$0.149605$&$0.670408$ \\
$0.32$&$0.0742308$&$0.150767$&$0.671556$\\
\hline
\end{tabular}
\end{center}\label{f8}

Now the surface redshift for a compact star model is obtained as, $z_s(R)=\frac{1}{\sqrt{\left(1-2\mathcal{U}\right)}}-1$. The numerical values of the surface redshift for different values of $\gamma$ have been presented in table.~1.
The gravitational redshift of our present model is calculated as,
\[z=e^{-\frac{\nu}{2}}-1,\Rightarrow z=\frac{1}{C}e^{-\frac{Br^2}{2}}-1,\] and its central value is obtained as $z_c=\frac{1}{C}-1$. Now $z_c>0$ gives, $1/C>1$ which consequently gives, $C<1$.
Now $\frac{dz}{dr}=-\frac{B}{C}e^{-\frac{Br^2}{2}}r$, at the origin $\frac{dz}{dr}=0$ and $\frac{d^2z}{dr^2}=-\frac{B}{C}<0$. It indicates that gravitational redshift is monotonic decreasing function of radius of the star.

\item {\bf Generating Function :} Based on the choice of a single monotone function subject to boundary conditions which generates all regular static spherically symmetric perfect-fluid solutions of Einstein equations, an algorithm was proposed by Lake \cite{lake2}. Herrera et al. \cite{h11} extended this work to the case of locally anisotropic fluids and proved that two functions instead of one is required to generate all possible solutions for anisotropic fluid. Thus they proved that any solution describing a static anisotropic fluid
distribution is fully determined by means of the two generating functions $\Pi$ and $Z$, where the expression for these two generating functions are given by,
\begin{eqnarray*}
\Pi(r)&=& 8\pi(p_r-p_t),\\
e^{\nu(r)}&=&e^{\int\left(2Z(\tau)-\frac{2}{\tau}\right)d\tau}.
\end{eqnarray*}
For our present model, these two generating functions are obtained as,
\begin{eqnarray}
\Pi(r)&=& 8\pi(p_r-p_t)=-8\pi \Delta,\label{st}\\
Z(r)&=&Br+\frac{1}{r}.
\end{eqnarray}
\end{itemize}
Where the expression of $\Delta$ present in eqn. (\ref{st}) has been given in eqn. (\ref{del5}).

\begin{table*}[t]
\centering
\caption{The numerical values of central density, surface density, central pressure, relativistic adiabatic index, mass and effective surface redshift for different values of $\gamma$ for the compact star SAX J1808.4-3658 by assuming $M = 0.88~M_{\odot},\, R = 8.9 $ km., Q = 0.0089.}\label{tbl2}
\begin{tabular}{@{}ccccccccccccc@{}}
\hline
$\gamma$&$\rho_c$&$\rho_s$&$p_c$&$\Gamma_{r0}$ & $M$ & $z_s(R)$\\
\hline
0.0&  $6.83316 \times 10^{14}$& $5.2929\times 10^{14}$ & $4.62079\times 10^{34}$ &4.76969& 1.298 & 0.188193\\
0.08&  $ 6.77128\times 10^{14}$ &$5.2744\times 10^{14}$ & $4.49062\times 10^{34}$ &4.85695&1.30959 &0.190383 \\
0.16& $ 6.71063\times 10^{14}$&  $ 5.25577\times 10^{14}$ & $4.36458\times 10^{34}$ &4.9459& 1.32074&0.192503\\
0.24&  $6.65118\times 10^{14}$&  $5.23702\times 10^{14}$ & $4.2425\times 10^{34}$ &5.03659& 1.33148& 0.194554\\
0.32& $6.5929\times 10^{14}$&  $5.21816 \times 10^{14}$& $4.12421\times 10^{34}$ &5.12908& 1.34183 & 0.19654\\
\hline
\end{tabular}
\end{table*}





\begin{figure*}[htbp]
    \centering
		\includegraphics[scale=.42]{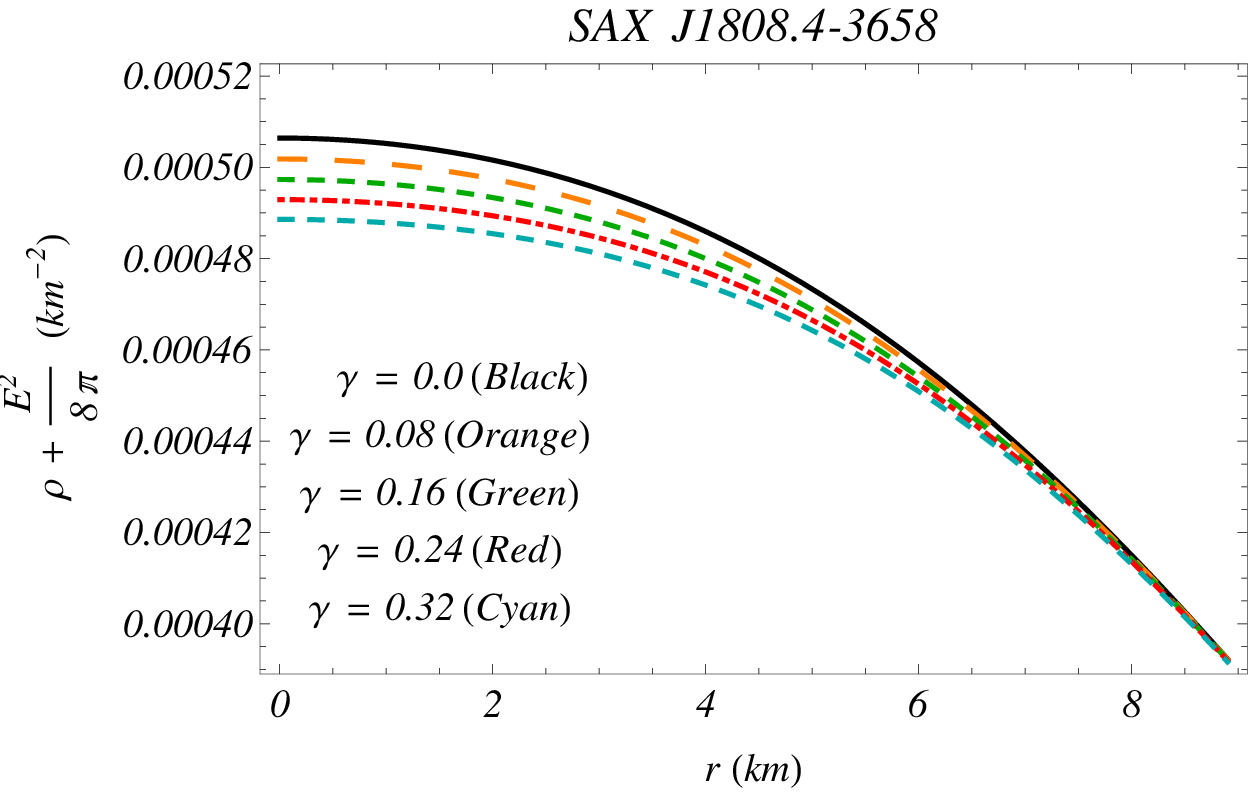}
        \includegraphics[scale=.42]{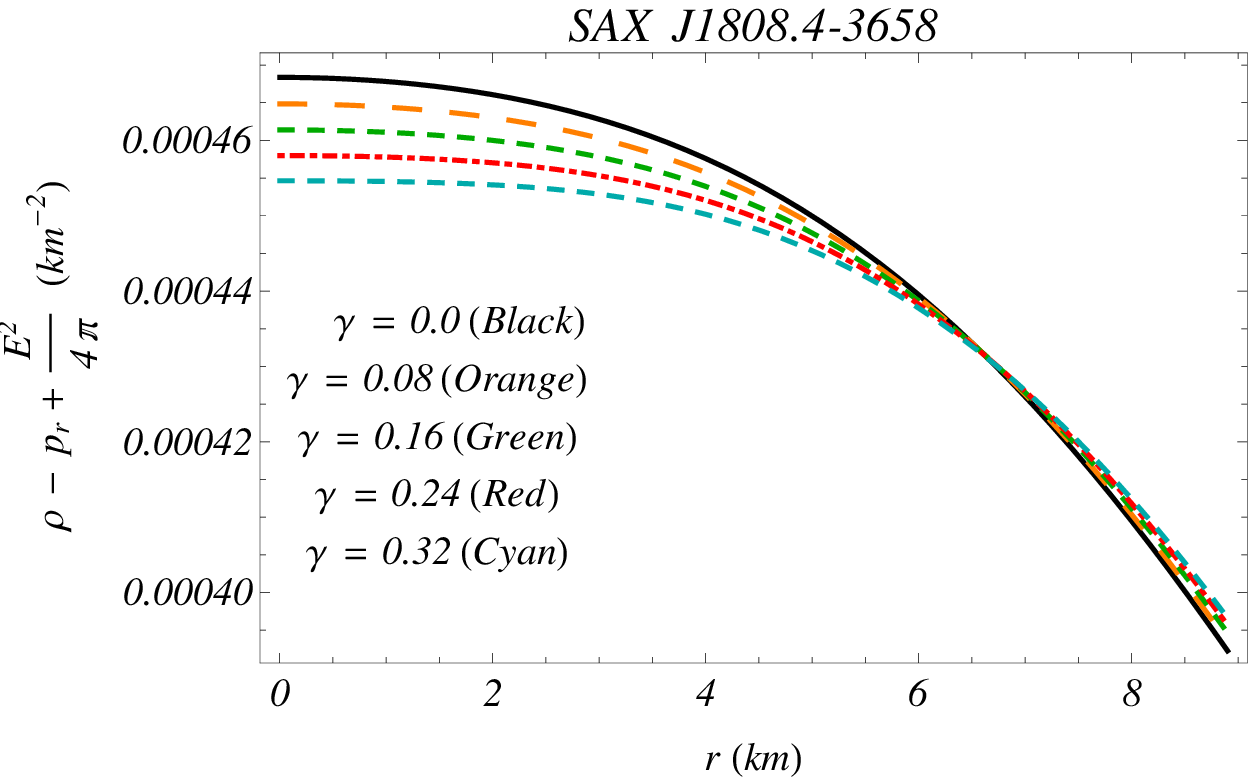}
        \includegraphics[scale=.42]{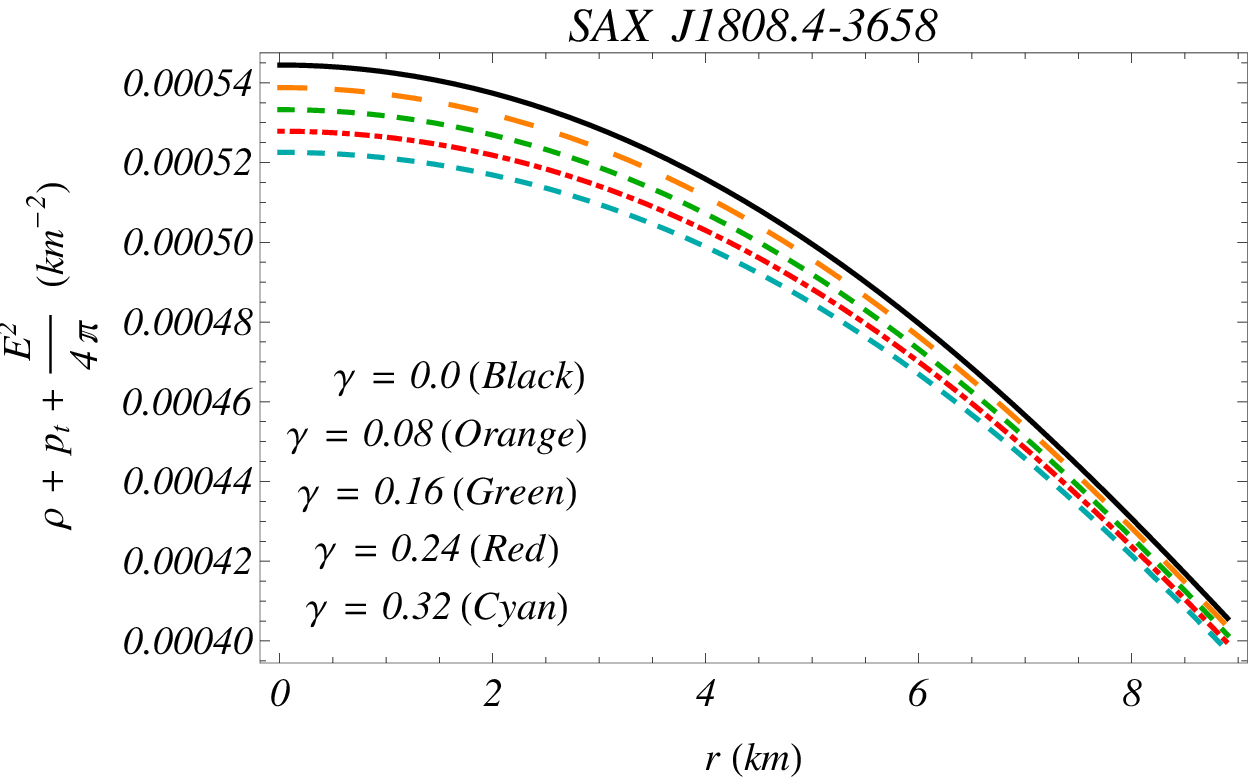}
        \includegraphics[scale=.4]{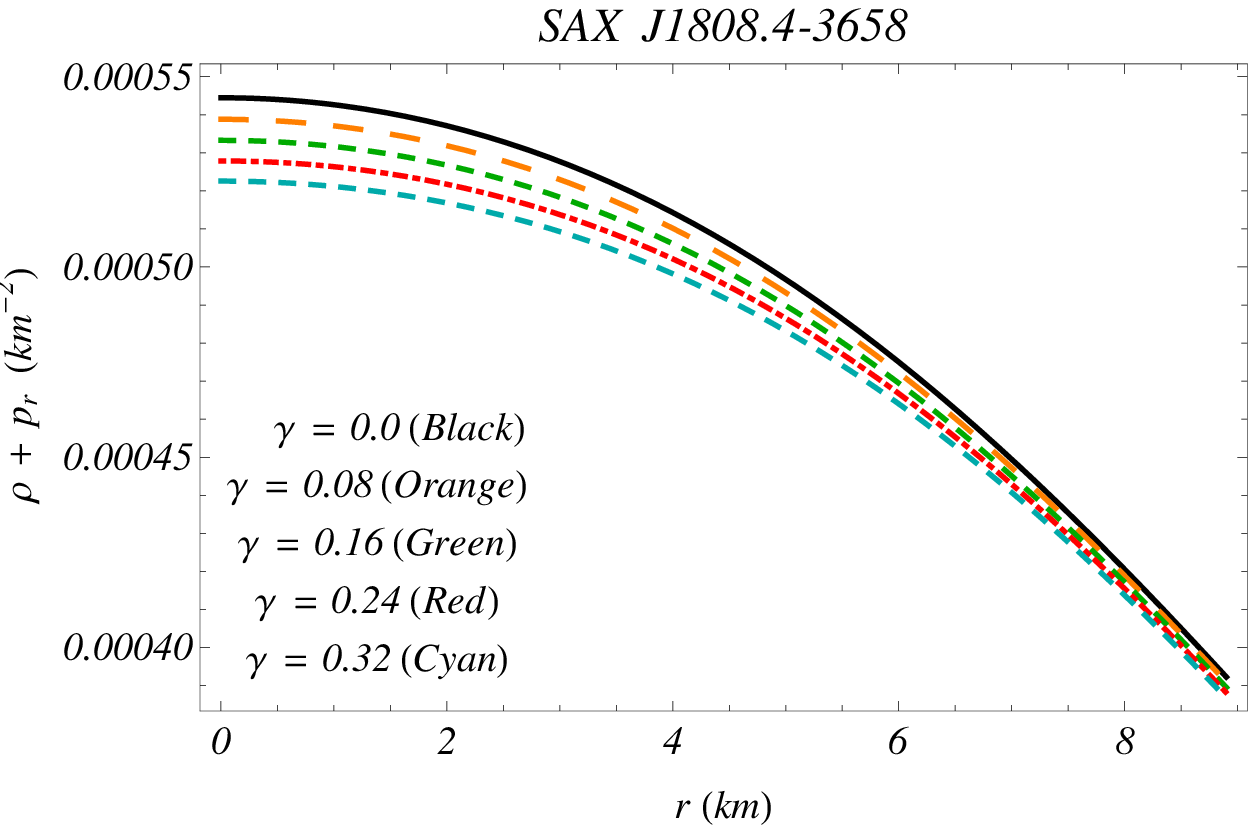}
        \includegraphics[scale=.42]{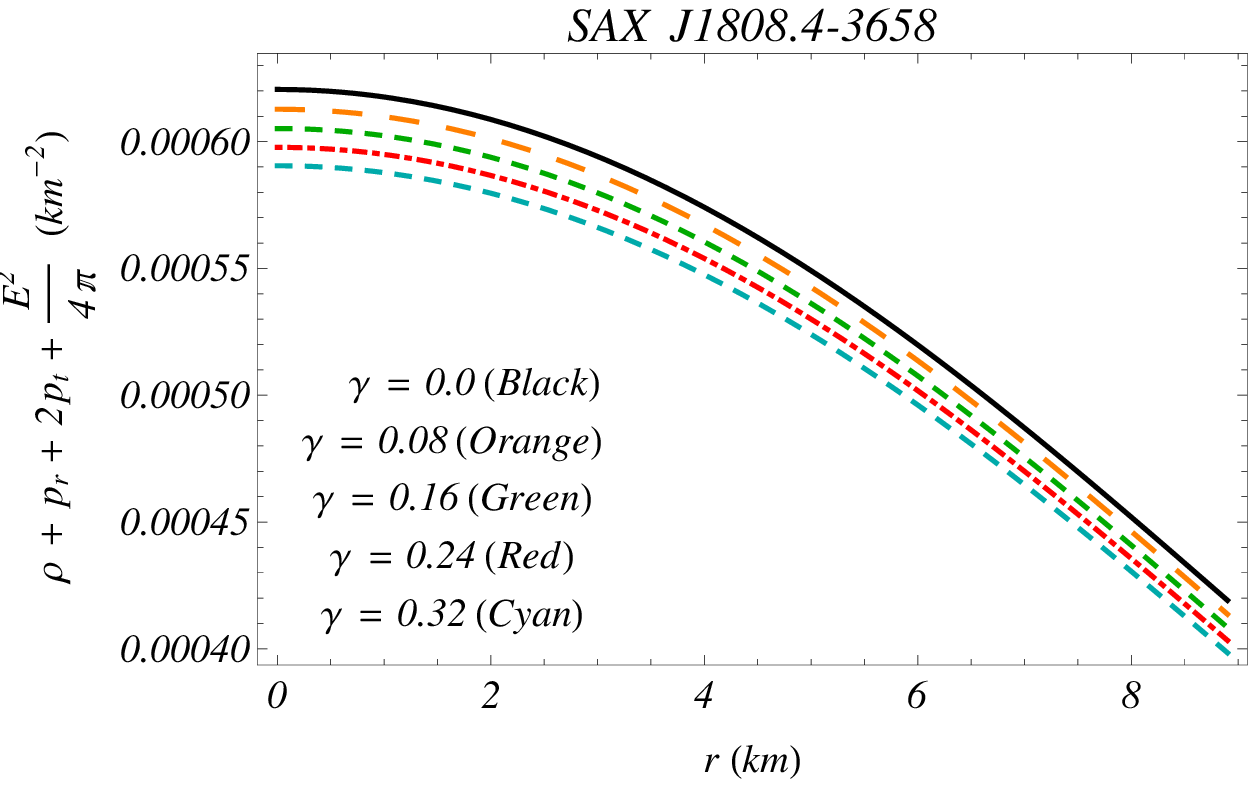}
        \includegraphics[scale=.42]{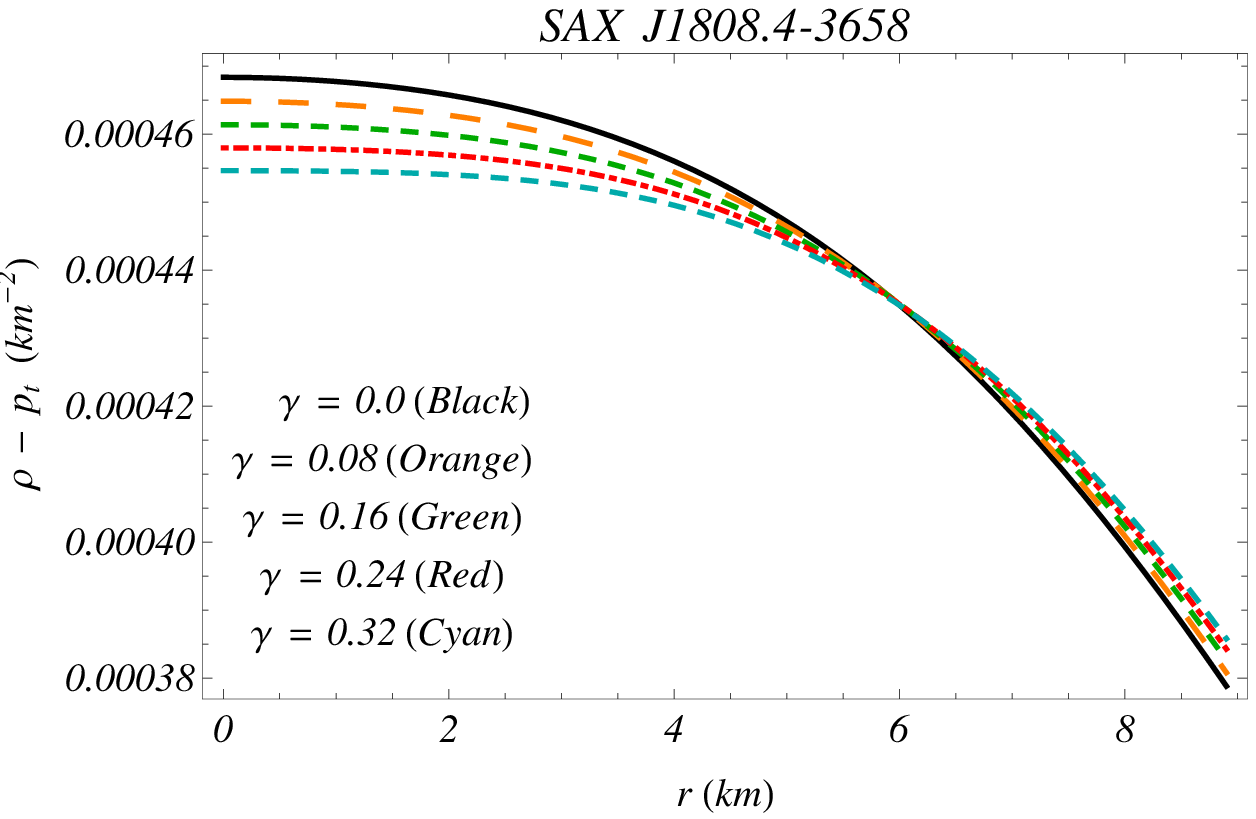}
         \includegraphics[scale=.42]{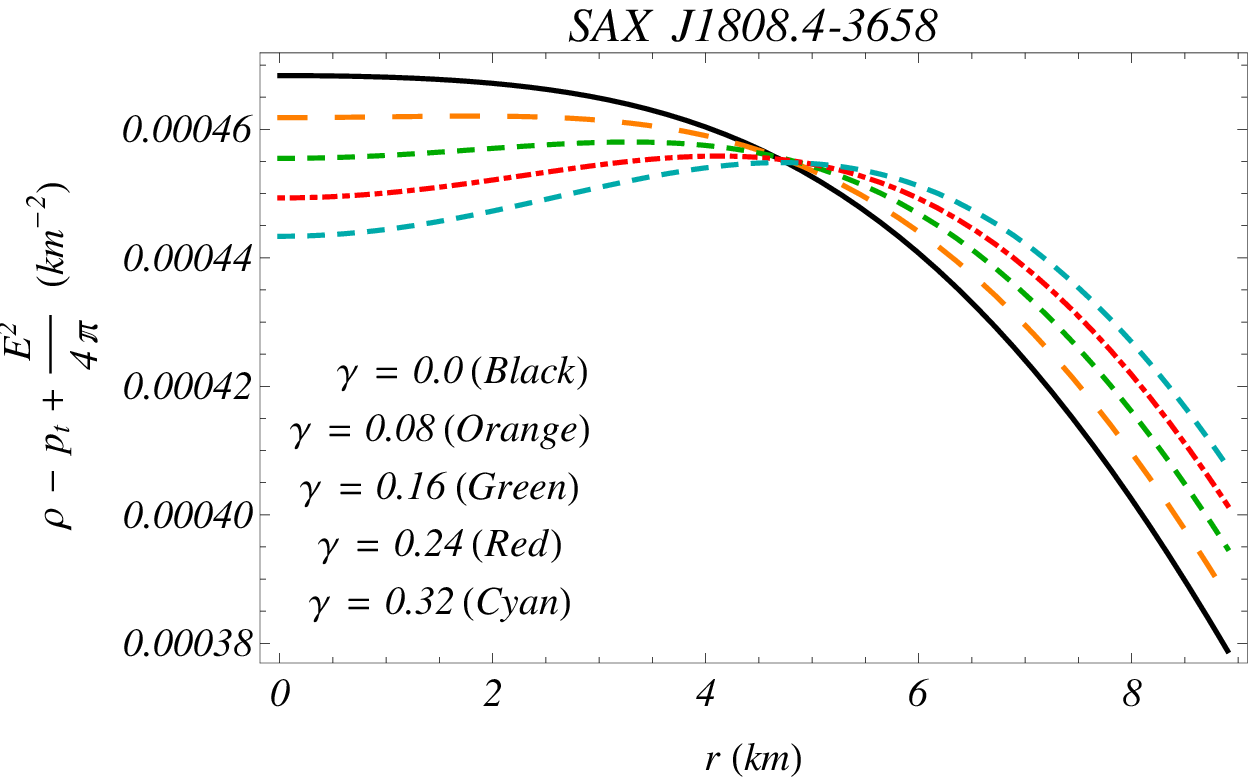}
       \caption{All the energy conditions are plotted inside the stellar interior for the strange star $SAX J1808.4-3658$ for different values of $\gamma$ mentioned in the figure.\label{ec9}}
\end{figure*}

\section{Equilibrium condition under different forces }\label{sec8}
In this section we shall check the equilibrium condition of our present model under different forces acting on our present system. The equilibrium equation can be spread into five different
forces namely: the hydrostatic force $F_h$, the gravitational force $F_g$, the anisotropic force $F_a$, the electric force $F_e$ and finally the force related to modified gravity, i.e., $F_m$. Moreover, the explicit form of these forces can be
written as :
\begin{figure}[htbp]
    \centering
        \includegraphics[scale=.45]{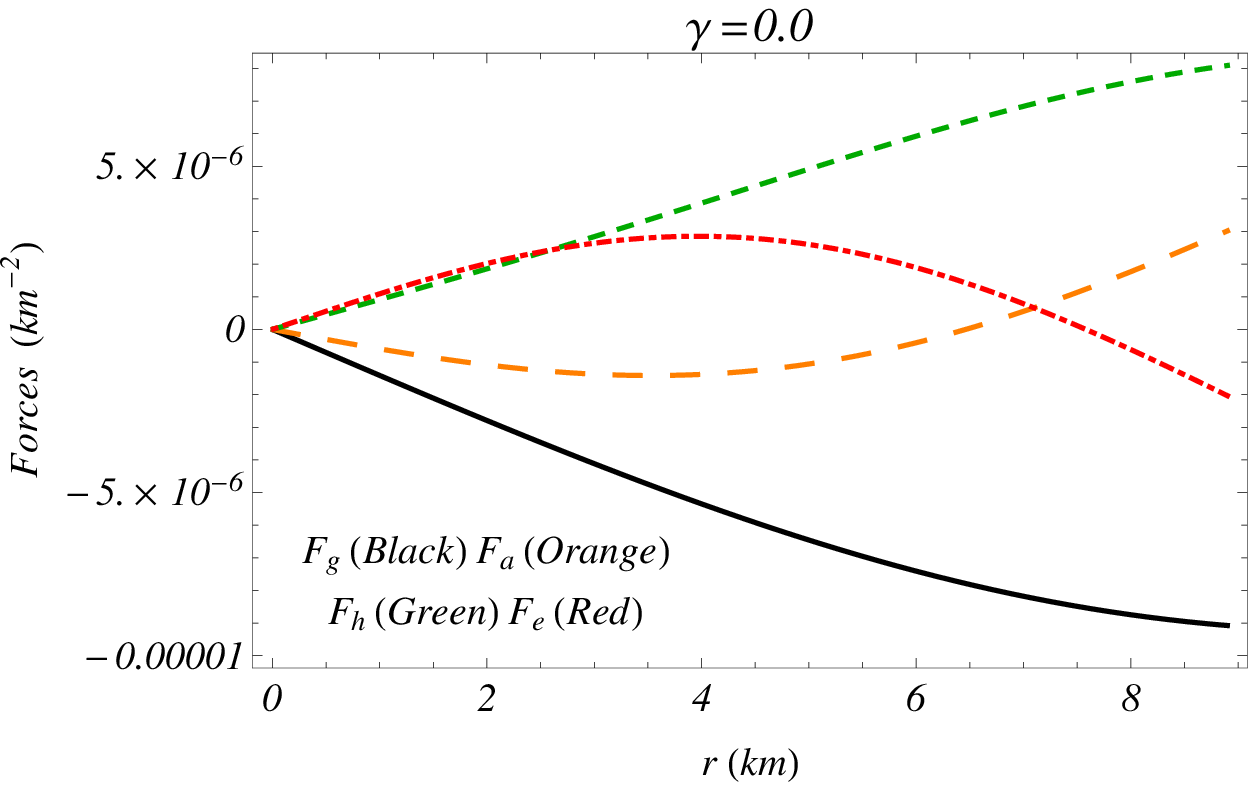}
        \includegraphics[scale=.45]{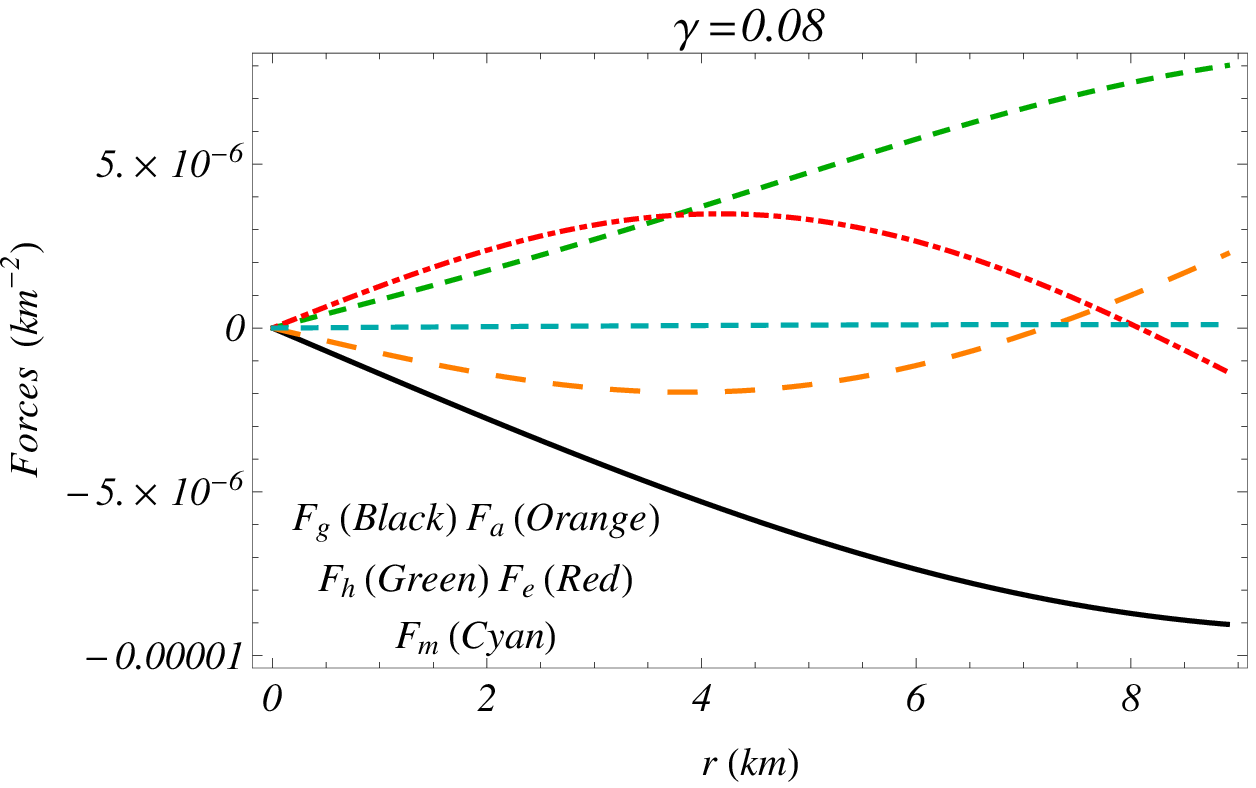}
        \includegraphics[scale=.45]{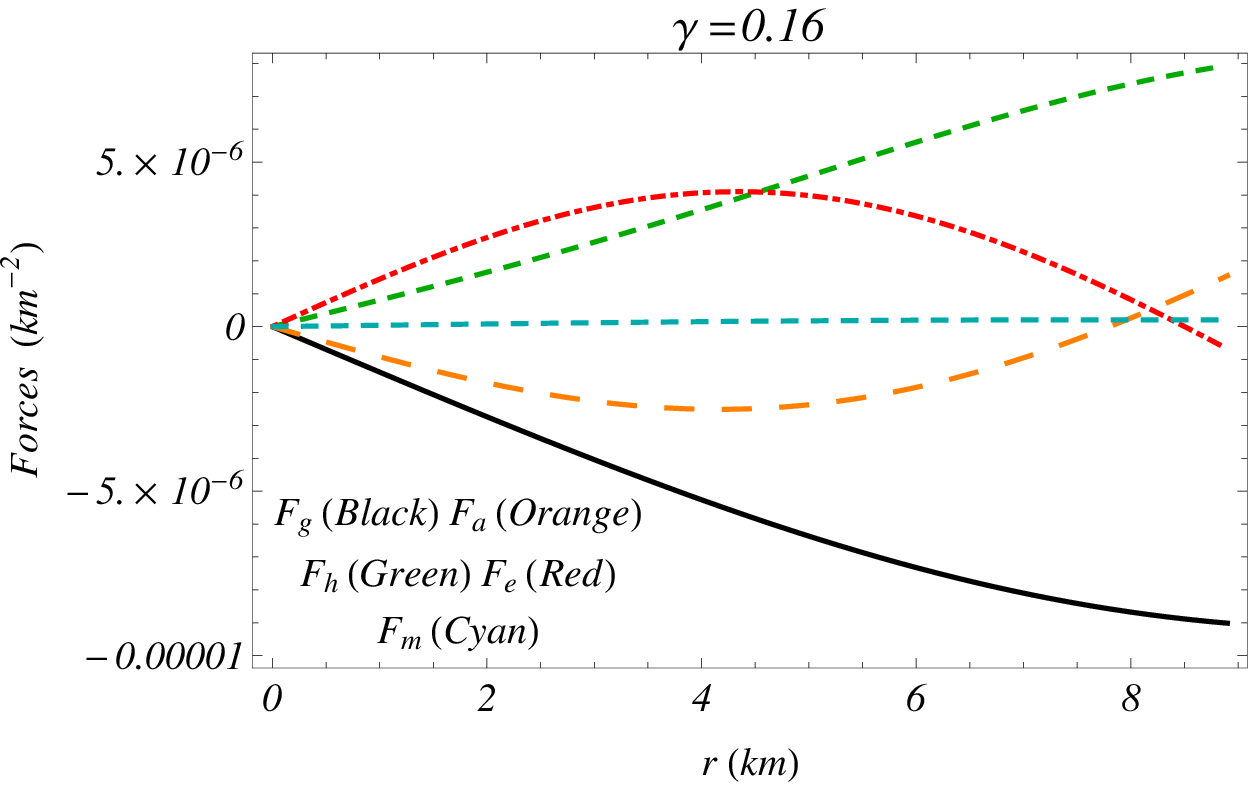}
        \includegraphics[scale=.45]{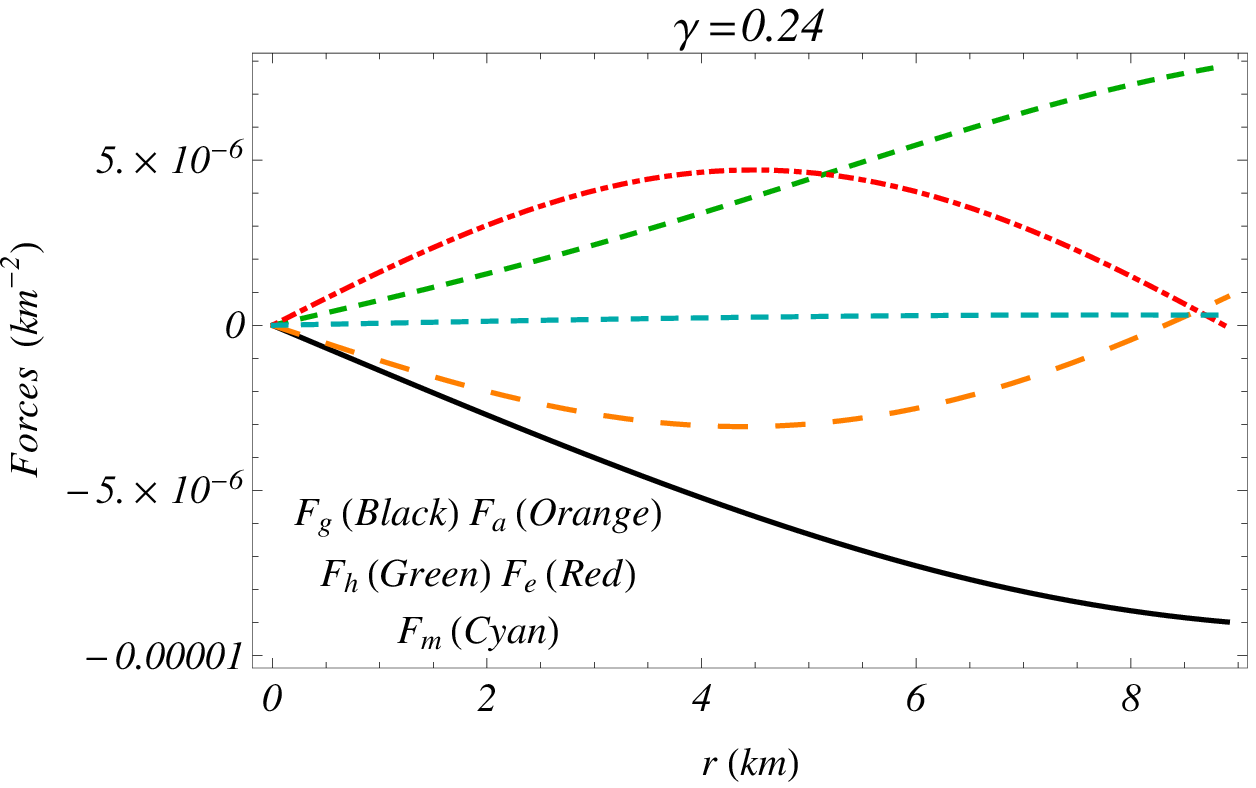}
        \includegraphics[scale=.45]{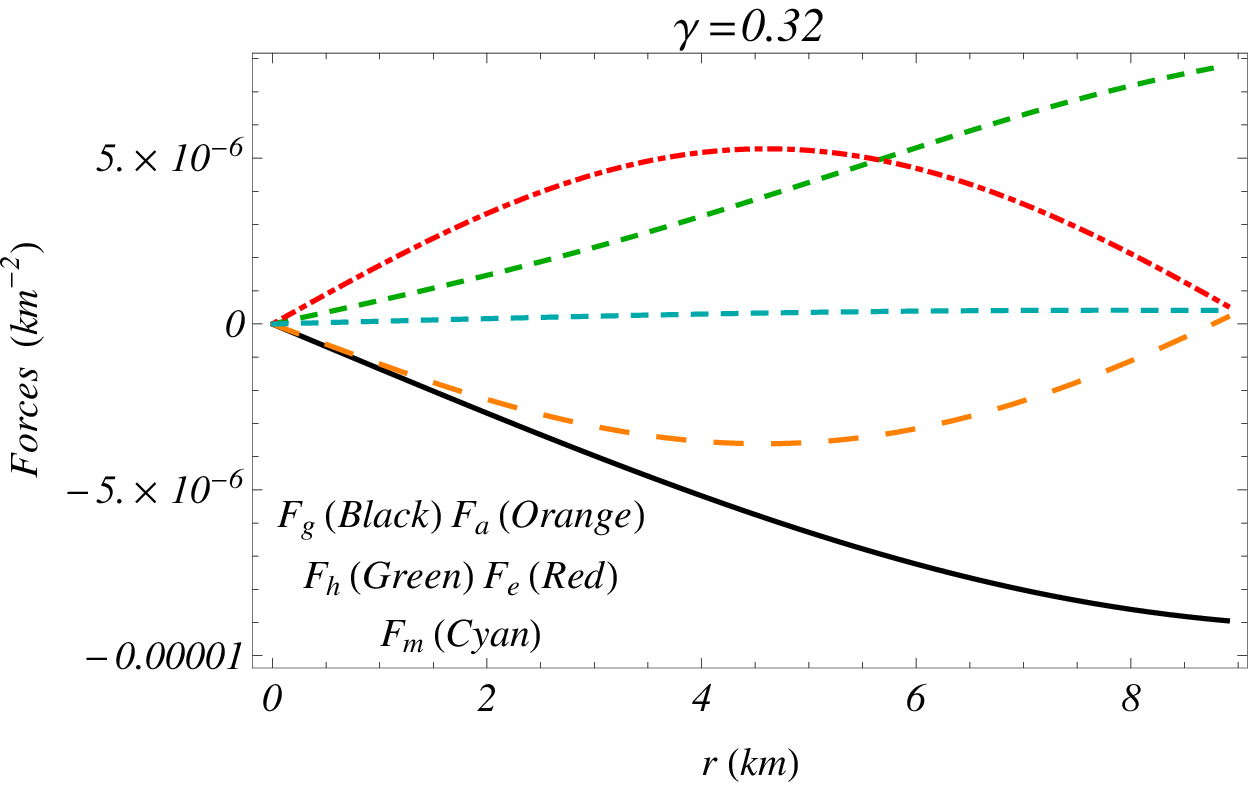}
       \caption{Different forces acting on the present model are plotted against r for the strange star candidate SAX J1808.4-3658 by taking different values of $\gamma$.\label{tov1}}
\end{figure}


\begin{eqnarray}
F_g&=&-\frac{\nu'}{2}(\rho+p_r)\nonumber\\
&=&-\frac{B r \big(a + B + (2 b + a B) r^2 + b B r^4 \big)}{(\gamma + 4 \pi){\Psi}^2},\\
F_h&=&-\frac{dp_r}{dr} \nonumber\\
&=& \frac{2 (a^2 - 4 b) r + \Psi r \big(a B + 2 b (3 + B r^2)\big)}{2 (\gamma + 4 \pi){\Psi}^3},\nonumber\\
F_a&=&\frac{2}{r}(p_t-p_r)=\frac{2}{r}\Delta \\
F_e &=&  \frac{8\pi}{8\pi+2\gamma}\frac{q}{4\pi r^4}\frac{dq}{dr} \nonumber \\
&=&\frac{1}{4\pi+\gamma}\left(\frac{2}{r}E^2+\frac{1}{2}\frac{d}{dr}(E^2)\right),\\
F_m&=& -\frac{\gamma}{8\pi+2\gamma}(\rho'+p_r'+2p_t')\nonumber\\
&=& \frac{\gamma r}{2 (\gamma + 4 \pi) (3 \gamma + 4 \pi) {\Psi}^3} \Big[4(b-a^2)+ 11 a B \nonumber\\
&&- 2 B^2-\big(2 a^3-7 a^2 B-26 b B + 2 a (4 b + B^2)\big) r^2   \nonumber\\
&& -3 b \big(2 (a^2 + b) - 7 a B\big) r^4 +2 b^2 (-b + B^2) r^8 \nonumber\\
&& +            2 b \big(5 b B + a (-3 b + B^2)\big) r^6 \Big] .
\end{eqnarray}
The TOV equation in modified gravity can be written as,
\begin{eqnarray}\label{con}
-\frac{\nu'}{2}(\rho+p_r)-\frac{dp_r}{dr}+\frac{2}{r}(p_t-p_r)\nonumber\\+\frac{\gamma}{8\pi+2\gamma}\left(\frac{d\rho}{dr}+\frac{p_r}{dr}+2\frac{dp_t}{dr}\right)
+\frac{8\pi}{8\pi+2\gamma}\frac{q}{4\pi r^4}\frac{dq}{dr}=0,\nonumber\\
\end{eqnarray}
In eqn. (\ref{con}), for $\gamma=0$ we regain the conservation equation in Einstein-Maxwell gravity. Now the above equation can be denoted by,
\begin{eqnarray}
F_g+F_h+F_a+F_m+F_e &=& 0.
\end{eqnarray}
The expression of different forces acting on our system are depicted in Fig.~\ref{tov1} for different values of the coupling constant $\gamma$.


\begin{figure}[htbp]
    \centering
        \includegraphics[scale=.48]{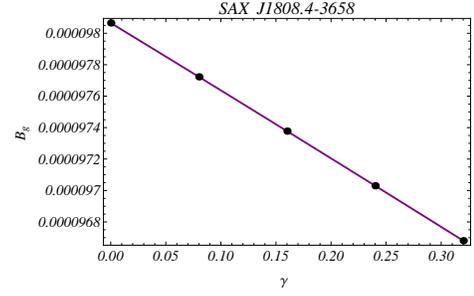}
       \caption{The variation of $B_g$ with respect to $\gamma$}\label{fgh1}
\end{figure}

\section{Discussion}\label{dis}

In this exposition we have successfully modeled the compact star SAX J 1808.4-3658 within the framework of $f(R,\,T)$ modified gravity theory. We have employed the physically motivated Tolman-Kuchowicz ansatz \cite{tk1,tk2} for the metric potentials. We also argued that most researchers consider an {\em ad hoc} equation
of state (EoS) $\rho + p = 0$ in constructing the electromagnetic mass model which
results a negative pressure. Instead of choosing this type of EoS, in order to solve the system of governing equations, we adopted the MIT Bag model equation of state. Our model has been subjected to rigorous regularity, causality and stability tests which highlighted the role of charge and the $f(R,\,T)$ coupling constant $\gamma$. The free constants arising from integrating the field equations are fixed through the boundary conditions. In order to bring out the contributions from the modified theory we have plotted the thermodynamical and physical properties of the star for various values of the $f(R,\,T)$ coupling constant. The main findings
of this analysis can be summarized as follows:
\begin{itemize}
  \item In fig.~\ref{metric} we have plotted the metric potential with respect to radius. The exterior spacetime is also shown in the figure. One can also note that, at the boundary, the interior and exterior metric coincides. The interior metric potentials are free from singularities and continuous inside the boundary. We can see that the metric potentials do not depend on the coupling constant $\gamma$.
  \item In fig.~\ref{pr8} the profiles of density and both radial and transverse pressures are plotted with respect to radius for different values of $\gamma$. The black, red, blue, purple and cyan colored plots correspond to $\gamma=0,\,0.08,\,0.16,\,0.24$ and $0.32$ respectively. The pressure and density all are monotonic decreasing functions of $r$, i.e., they have maximum value at the center and then it gradually decrease towards the boundary. It is also verified that density and both the pressures are non negative inside the stellar interior and at the boundary $r=R$, $\rho(r=R)>0$, $p_t(r=R)>0$ for different values of $\gamma$. The radial pressure $p_r$ vanishes at the point $r=R$, i.e., it determines the size of the compact object.
  \item In fig.~\ref{sv} we have shown the behavior of the radial and transverse velocity of sound and the stability factor for different values of $\gamma$. One can note that the radial velocity of sound is independent of $\gamma$ but the transverse velocity of sound depends on coupling constant. For higher values of $\gamma$ the value of $V_t^2$ increases. The profiles of $|V_t^2-V_r^2|$ have been plotted for different values of $\gamma$ and it is clear that Andr\'{e}asson condition is satisfied.
      \item In fig.~\ref{gam1} the profiles of radial relativistic adiabatic index have been shown for $\gamma=0,\,0.08,\,0.16,\,0.24$ and $0.32$. For larger values of $\gamma$, the $\Gamma_r$ takes higher values. $\Gamma_r$ is monotonic increasing function of `r' and greater than $4/3$ everywhere inside the fluid sphere for each values of $\gamma$ mentioned in the figure and hence our model of charged compact star is stable.
          \item In fig.~\ref{ec3} the nature of the electric field and pressure anisotropy have been plotted for different values of $\gamma$ and one can note that $E^2$ is positive everywhere inside the fluid sphere. Moreover we note that $E^2$ is monotonic increasing upto about $7$ km. for different values of $\gamma$ then it becomes monotonic decreasing.
\end{itemize}
We found that an increase in the coupling constant lowers the density, radial and transverse pressures at each interior point of the configuration while there is a corresponding increase in the star's mass. Further investigation by Astashenok et al. \cite{ast}
gives a non perturbative model of strange spherical objects
in $f(R)=R+2\alpha R^2$ gravity theory, where $\alpha$ is a constant.
They have shown that the masses of candidate strange
spherical objects increase when the value of the parameter $\alpha$
increases progressively. In our present work, we have also obtained the same trend. We conclude that $f(R,\,T)$ stars have larger masses compared to their 4D classical GR counterparts. The stability of the star is derived from the fact that the force due to anisotropy is attractive ($p_t < p_r$) up to some radius $r = r_0$ within the boundary. The repulsive contributions from the electromagnetic field and the $f(R,\,T)$ coupling constant help stabilize the inner core. The anisotropic factor becomes positive closer to the surface layers of the star leading to greater stability in this region. The surface redshift increases as $\gamma$ increases. It is well-known that presence of  anisotropy within the stellar core lead to higher surface redshifts compared to an isotropic, perfect fluid matter distribution.
\begin{itemize}
  \item In fig.~\ref{eos} the equation of state parameter $\omega_r$ and $\omega_t$ have been plotted for different values of the coupling constant and we note that both are monotonic decreasing function of `r' and moreover $0<\omega_r,\,\omega_t<1$ and it corresponds to radiating era \cite{sharif1}. The variation of radial pressure with respect to density follows a linear relationship which is clear from our assumption. On the other hand the variation of transverse pressure with respect to density follows almost a parabolic nature.
  \item The charged compact star model obeys all the energy conditions namely null energy conditions, weak energy conditions, strong energy conditions and dominant energy conditions for different values of $\gamma$ which have been shown in fig.~\ref{ec9}.
  \item In fig.~\ref{tov1} all five different forces namely gravitational force, anisotropic force, hydrostatics force, electric force and force due to modified gravity have been shown for $\gamma=0,\,0.08,\,0.16,\,0.24$ and $0.32$. We see that the gravitational force $F_g$ is always negative and hydrostatics force $F_h$ is always positive and the another three forces show mixed behavior. In this situation, the contribution of the force due to modified gravity $F_m$ is very negligible compared to other four forces so that it seems likely to be overlapped with the X-axis. This means that the effect of coupling also becomes less effective.
  \item In fig.~\ref{fgh1} we have shown the variation of the bag constant $B_g$ with respect to coupling constant $\gamma$ and it is observed that the value of the bag constant takes lower value when $\gamma$ increases.
\end{itemize}
To summarize, in the present paper we have obtained a singularity-free model of charged anisotropic compact star in presence of electric field in $f(R,\,T)$ modified theory of gravity and the result have been analyzed both analytically and graphically. We show that solution of the field equations depend on the MIT bag model EoS which is a familiar equation of state already used by several researchers
for modeling compact star. Finally, it is worth mentioning that taking $\gamma \rightarrow 0$ GR
results in four dimensions are recovered.

{\bf ACKNOWLEDGMENTS:} PB is thankful to IUCAA, Government of India for providing visiting associateship.

\end{document}